\documentclass[aps,pre,subeqn,nofootinbib,notitlepage,amsmath,amsfonts]{revtex4-1}
\usepackage{setspace}
\usepackage{graphicx}
\usepackage{subfigure}

\usepackage{GlavDesign}

\newcommand{\scaleprofile}{0.7}
\newcommand{\scalevmd}{0.46}
\newcommand{\co}{CO$_2$ }
\newcommand{\ch}{CH$_4$ }
\newcommand{\coeq}{CH_4}
\newcommand{\cheq}{CO_2}
\newcommand{\waeq}{H_2O}

\renewcommand{\eqr}[1]{\ref{#1}}
\renewcommand{\secr}[1]{\ref{#1}}
\renewcommand{\figr}[1]{\ref{#1}}

\newcommand{\dispersion}[2]{\langle #1 #2 \rangle - \langle #1 \rangle \langle #2 \rangle}
\newcommand{\fluctuation}[2]{\langle \delta #1 \, \delta #2 \rangle}

\begin{document}

\title{Adsorption of \co and \ch and their mixtures in gas hydrates.}

\author{K.~S.~Glavatskiy\tsup{1,2}}
\author{T.~J.~H.~Vlugt\tsup{2}}
\author{S.~Kjelstrup\tsup{1,2}}
\affiliation {
\tsup{1}Department of Chemistry, Norwegian University of Science and Technology, NO 7491 Trondheim, Norway.\\
\tsup{2}Department of Process and Energy, Delft University of Technology, Leeghwaterstraat 44, 2628 CA Delft, The Netherlands. }
\date\today

\begin{abstract}
We report results from grand-canonical Monte Carlo simulations of methane and carbon dioxide adsorption in structure sI gas hydrates. Simulations of pure component
systems show that all methane sites are equivalent, while carbon dioxide distinguishes between two types of sites, large or small. The adsorbed mixture can be
regarded as ideal, as long as only large sites are occupied. A strong preference is demonstrated for methane, when the smaller sites become filled.

The molar heat of adsorption of methane decreases with composition, while the molar heat of adsorption for carbon dioxide passes an extremum, essentially in
accordance with the observation on the site sizes. The Helmholtz energies of the hydrate with \co-\ch gas mixture for temperatures between 278 and 328 K and
pressures between 10$^4$ and 10$^9$ Pa indicate that certain mixtures are more stable than others. The results indicate that a thermodynamic path exists for
conversion of a pure methane hydrate into a pure carbon dioxide hydrate without destroying the hydrate structure.
\end{abstract}

\maketitle


\numberwithin{equation}{section}

\section{Introduction}\label{sec/Introduction}

Understanding the possibility to exchange \ch with \co in hydrates is of great importance to the oil industry, and has been the focus of several studies
\cite{Sloan2003a, SloanHydrates,Tegze2007,Baldwin2009, Kvamme2007a}. By studying the radial distribution function of a fully occupied sI hydrate using molecular
dynamic simulations, Geng et al \cite{Geng/COCH} have recently suggested that a \co+ \ch mixture in the hydrate lattice can be more stable than a hydrate with pure
components, \ch or CO$_2$. The present work aims to find conditions for which this is true, using Grand-Canonical Monte-Carlo simulations (GCMC) in the study of the
adsorption of pure \co and CH$_4$, as well as mixtures of \co and \ch in an sI hydrate.

Monte-Carlo simulations of this type are excellently suited for this purpose, because they allow one to study also metastable configurations, in particular,
non-fully occupied hydrates \cite{Papadimitriou2008a, Katsumasa2007, Wierzchowski2007}. Molecular simulations allows one to study the conditions of hydrate
formation which are complicated to work with experimentally. For a particular configuration of a hydrate with adsorbed gas(es) one can use the adsorption isotherms
to obtain the Helmholtz energy, which can be used to give information about relative stabilities. GCMC simulations have been used extensively to study adsorption
processes \cite{Vlugt2008, Karavias1991} and we shall take advantage of the method development described earlier \cite{Wierzchowski2007, Wierzchowski2006}.

We shall use the chlatrate sI structure, which forms at typical reservoir conditions (for example, pressures up to $10^4$ bar or $10^9$ Pa, which corresponds to
reservoir depths up to 10 km \cite{SloanHydrates}). The structure, which is well established \cite{McMullan1965a}, has 8 cages per unit cell. Interaction potentials
for water are available through a series of systematic investigations \cite{Castillo2009}. We will use a rigid water model and the rigid framework model for all of
the calculations. The typical behavior of the adsorption isotherms remain essentially unaltered when the hydrate framework model becomes mobile (and therefore is
not presented here).

The Helmholtz energy difference is negative for a spontaneous process at constant volume and temperature. The process of practical interest in this study can be
written schematically as the reaction
\begin{equation}\label{eq/Intro/01}
N_{\cheq} + N_{\coeq} + N_{\waeq} \rightleftharpoons (N_{\cheq} + N_{\coeq}) \cdot N_{\waeq}
\end{equation}
The Helmholtz energy difference per mole of formed unit cell in this reaction is denoted by $\Delta_r F$. The long-term aim of the work is to obtain the values for
$\Delta_r F$ which would decide on whether the hydrate formation is favorable or not. This paper concerns a first step in this direction. We will study the
variation in the Helmholtz energy of the hydrate with a mixed gas adsorbed, $(N_{\cheq} + N_{\coeq}) \cdot N_{\waeq}$, compared to a metastable hydrate with empty
cages $N_{\waeq}$. This difference is denoted by $\Delta F$ in this paper. Since, adding a gas molecule in a cage stabilizes the hydrate, this Helmholtz energy
difference will normally decrease with the loading. However, if a gas hydrate with two components adsorbed is less stable than a single component one, the variation
in the Helmholtz energy would also increase with loading. The main purpose of the paper is to show this, and discuss how this process can be understood in terms of
molecular behavior. The quantities $\Delta_r F$ and $\Delta F$ are related, but we return to the relationship and a discussion of $\Delta_r F$ in a subsequent
paper.

The paper is organized as follows. In \ref{sec:Thermodynamics} we discuss the basic thermodynamic relations which allow us to calculate the Helmholtz energy and the
molar heat of adsorption as a function of loading. The details of the simulations are specified in \secr{sec/Simulations}. In \secr{sec/Results} we provide our
results. We report adsorption isotherms for the single component hydrates and the mixture of \co and \ch in the hydrate, as well as the selectivity data for the
mixture adsorption. Furthermore, we calculate the Helmholtz energy difference $\Delta F$ of the above hydrates as a function of loading. Finally, we compare the
data for the molar heat of adsorption with the structure of the hydrates. An overall discussion and conclusion are given in \secr{sec/Conclusions}.

\section{Thermodynamics of adsorption}\label{sec:Thermodynamics}
\subsection{Helmholtz energy}\label{sec/FreeEnergy}

We consider the framework of water molecules in the sI clathrate structure. In simulations we can control whether this framework is flexible or rigid. In both cases
we specify and keep fixed the lattice parameters of the framework, so the volume of the system $V$ is constant. We also keep the temperature of the system $T$
fixed.

The Helmholtz energy $F$ of the clathrate with given numbers of adsorbed guest molecules (a loading) is given by
\begin{equation}\label{eq/FreeEnergy/01}
F = -p\,V + \mu_{w}\,N_{w} + \sum_{i=1}^{n}{\mu_{i}\,N_{i}}
\end{equation}
Here $p$ is the pressure in the system, $\mu_{i}$ is the chemical potential of the $i$-th guest component and $N_{i}$ is the number of adsorbed molecules of
component $i$, while $\mu_{w}$ and $N_{w}$ are the corresponding quantities for water. The number of compounds is $n$. For pure components, $n=1$, for a mixture of
methane and carbon dioxide $n=2$. The number of water molecules is fixed, while the number of guest molecules vary.  The chemical potentials, and therefore the
Helmholtz energy, depend on the loading.

We would like to calculate $\Delta F(N_{\{i\}}) \equiv F(N_{\{i\}}) - F(0)$, the difference in the Helmholtz energy of a filled clathrate $F(N_{\{i\}}) \equiv
F(N_{w}, N_{1}, \cdots, N_{n})$ and the Helmholtz energy of the empty clathrate $F(0) = F(N_{w}, 0, \cdots, 0) =  -p\,V + \mu_{w}\,N_{w}$. We refer to this
difference as the \textit{Helmholtz energy of the mixture of gases in the hydrate}. It follows from \eqr{eq/FreeEnergy/01} that
\begin{equation}\label{eq/FreeEnergy/03}
\Delta F(N_{\{i\}}) = -\left[p(N_{\{i\}})-p(0)\right]V + \left[\mu_{w}(N_{\{i\}})-\mu_{w}(0)\right]N_{w} + \sum_{i=1}^{n}{\mu_{i}(N_{\{i\}})\,N_{i}}
\end{equation}
where $p(N_{\{i\}})$ is the hydrostatic pressure at the given volume and loading, while $p(0)$ is the hydrostatic pressure at the same volume and zero loading.
According to the Gibbs-Duhem relation we have
\begin{equation}\label{eq/FreeEnergy/04}
S\,dT - V\,dp + N_{w}\,d\mu_{w} + \sum_{i=1}^{n}{N_{i}d\mu_{i}} = 0
\end{equation}
Integrating the equation at a constant temperature, volume, and a constant number of water molecules from zero loading to a given loading we obtain
\begin{equation}\label{eq/FreeEnergy/05}
- V\left[p(N_{\{i\}})-p(0)\right] + N_{w}\left[\mu_{w}(N_{\{i\}})-\mu_{w}(0)\right] + \sum_{i=1}^{n}{\int_{0}^{\mu_{i}^*(N_{\{i\}})}N_{i}\,d\mu_{i}(N_{\{i\}})} = 0
\end{equation}
which, after substitution into \eqr{eq/FreeEnergy/03} and integration by parts, gives \cite{Wierzchowski2007}
\begin{equation}\label{eq/FreeEnergy/06}
\Delta F(N_{\{i\}}) = \sum_{i=1}^{n}{\mu_{i}(N_{\{i\}})\,N_{i}} - \sum_{i=1}^{n}{\int_{0}^{\mu_{i}^*(N_{\{i\}})}N_{i}\,d\mu_{i}(N_{\{i\}})}
\end{equation}
where $\mu_{i}^*$ is the specified chemical potential of each of the component.  The right hand side of this expression can be easily obtained from GCMC
simulations, in which one specifies the chemical potential of the adsorbing components and measures the number of adsorbed molecules. Furthermore, this form is
convenient for integration, because at low loadings the slope of the function $N_i(\mu_i)$ approaches zero. The integration in \eqr{eq/FreeEnergy/06} over the
chemical potential of one component is done by keeping the chemical potentials of all other components fixed. For a two-component system this means that integration
is performed first along the path of $\mu_2 = -\infty$ from $\mu_1 = -\infty$ to $\mu_1 = \mu_1^*$ and then along the path $\mu_1 = \mu_1^*$ from $\mu_2 = -\infty$
to $\mu_2 = \mu_2^*$.

The chemical potential of the $i$-th component directly follows from its fugacity $f_{i}$ \cite{ll5, moranshapiro}:
\begin{equation}\label{eq/FreeEnergy/07}
\mu_{i} = -k_{B}\,T\,\ln(R\,q_{i}) + k_{B}T\ln(f_{i})
\end{equation}
where the first term in \eqr{eq/FreeEnergy/07} is the contribution from the ideal gas. Here $q_{i} \equiv Z_{\rm{int},i}/(N_{A}\Lambda_{i}^{3})$ where
$Z_{\rm{int},i}$ is the partition function for internal degrees of freedom. If the molecules are rigid (which is the case for our simulations, see
\secr{sec/Simulations}), $Z_{\rm{int},i}$ has contributions only from rotation of the molecule. At high temperature  $Z_{\rm{int},i}= T/T_{r}$, where $T_{r}$ is the
characteristic temperature of the rotational degrees of freedom. Furthermore, $\Lambda_{i} \equiv hN_{A}/\sqrt{2\pi M_{i} R T}$ is the thermal de Brogile wavelength
of component $i$, $M_{i}$ is the molar mass, and $h$, $N_{A}$, $R$, $k_{B}$ are Plank's constant, Avogadro's number, the universal gas constant, and Boltzman's
constant, respectively.

\subsection{The heat of adsorption}\label{sec/HeatOfAdsorption}

The heat of adsorption of gas(es) into the hydrate at constant volume is  equal to the internal energy change of the reaction \eqref{eq/Intro/01} at constant
volume. From the internal energy of reaction \eqref{eq/Intro/01}, we define the partial molar energy for adsorption of a component $i$;
\begin{equation}
\Delta_r U_i = \left( \frac{\partial \Delta_r U}{\partial N_i} \right)_{N_j,V,T} = \left( \frac{\partial (U^h -U^g) }{\partial N_i} \right)_{N_j,V,T}
\end{equation}
where $U^{h}$ is the internal energy of the hydrate phase and $U^g$ is the corresponding value for the gas phase. For a guest molecule in the hydrate, both energies
have constant contributions from the translational energy, ${s_{i} \over 2} k_{B} T N_{i}$, where $s_{i}$ is the number of degrees of freedom of the component, and
these contributions cancel in the difference \cite{AdsorptionSimulations}. The remaining contribution is due to the difference between the configurational parts of
the internal energy $U^{h}_{c}$ and $U^{g}_{c}$. In the literature it is common to consider the gas to be ideal, so that it acts as a reference state to an adsorbed
phase \cite{Vlugt2008, AdsorptionSimulations}. Therefore, $U^{g}_{i,c} = N_{i}RT$. This defines the partial molar heat of adsorption $q$ as

\begin{equation}\label{eq/HeatOfAdsorption/03}
- q_i = \left(\frac{\partial U^{h}_{c}}{\partial N_i}\right)_{N_j,T, V} - RT
\end{equation}

In this study we are interested only in the hydrate phase. The quantity $-q_i$ defined by \eqr{eq/HeatOfAdsorption/03} is then a property of a hydrate phase with
reference to an ideal gas phase. This is typical in the simulation community and conforms with the discussion in \secr{sec/FreeEnergy}, to describe the hydrate
phase only \cite{Vlugt2008}. We shall use \eqr{eq/HeatOfAdsorption/03} for hydrates filled with single components, and calculate the molar heat of adsorption of \ch
and CO$_2$.

The derivative of the internal energy of the hydrate with respect to the number of adsorbed molecules can be calculated in a number of ways \cite{Vlugt2008}. In
GCMC simulations, one calculates the molar heat of adsorption \eqref{eq/HeatOfAdsorption/03} from the fluctuations of the thermodynamic quantities as
\cite{AdsorptionSimulations, Karavias1991}
\begin{equation}\label{eq/HeatOfAdsorption/06}
q = RT - \frac{\fluctuation{U^{h}_{c}}{N}}{\fluctuation{N}{N}}
\end{equation}
where $\fluctuation{X}{Y} \equiv \dispersion{X}{Y}$, and $\langle \rangle$ denotes an ensemble average in the grand-canonical ensemble.

\section{Simulation details}\label{sec/Simulations}

We perform grand-canonical Monte Carlo ($\mu$VT) simulations of \ch and \co in a 2x2x2 unit cell of sI hydrate, which has 64 cages, with a lattice parameter 12.03
{\AA} \cite{McMullan1965a}. The number of running cycles was varying from 500 to 500000 in order to achieve the desired accuracy. The system became equilibrated
rather quickly, typically after 300-500 cycles. The number of MC moves per cycle is equal to the number of particles of each component in the system, with a minimum
of 20.

In GCMC simulations one specifies the chemical potential of a component and calculates the average number of particles which correspond to this chemical potential.
For this a number of Monte Carlo moves is performed, typically attempts to displace a particle from one position in the box to another and to exchange particles
with the reservoir. In the semi-grand canonical MC simulations, we specify $N_{w}$ and $f_i$ or $\mu_{i}$, $i=1..n$ and we find the loading of the guest components
$N_{i}$ and accordingly the Helmholtz energy of the gas mixture in the hydrate. In our GCMC simulations the molecules were allowed to change their position and
orientation. They were also subjected to Regrow, Swap, and Identity change \cite{Martin1997} MC moves. Furthermore, a series of NVT simulations has been performed
to analyze the distribution of molecules between the cages. In this case the number of adsorbed molecules was specified, while the chemical potential was adjusted
during the simulation. The guest molecules were allowed to change the position and orientation as well as regrow.

The oxygen positions of water molecules were taken from crystallographic data \cite{McMullan1965a}. The orientation of hydrogen atoms was chosen random in
accordance with the Bernal-Fowler rule \cite{BernalFowler}. Initially all the hydrogen atoms were assigned randomly to each of 4 sites of the water molecule (2
sites from the hydrogens and 2 sites for the hydrogen bonds to the neighboring molecule). Then a short Monte Carlo procedure has been performed to displace the
hydrogen atoms between the sites randomly, in order to satisfy 2 conditions: i) each oxygen atom has only 2 hydrogen atoms, and ii) there is only 1 hydrogen atom
between each 2 oxygen atoms. The third requirement, that the number of the hydrogen atoms of the corresponding Wyckoff type is given, was satisfied automatically
after the above procedure.

The water model is TIP5PEw \cite{TIP5PEW}. We performed simulations with both immobile and mobile water molecules, so in the latter case they were allowed to change
their position and orientation. During the simulations we have found that behavior of the system with mobile and immobile water molecules is similar, while the
fluctuations were larger for the case of mobile molecules. In the analysis of the results we therefore present the data for a rigid water framework. The description
of the guest molecules was taken from the TraPPE force field \cite{TRAPPE}.

We performed simulations for the range of the temperature from 278 K to 328 K. The adsorption isotherms were qualitatively the same for all the temperatures in this
range. We therefore focused on the temperature of 278 K.

The adsorbed components were assumed to be in equilibrium with an imaginary gas reservoir. Specifying the chemical potentials of the components in this gas
determines therefore the chemical potentials of the components, adsorbed into the hydrate. This procedure is typical in the molecular simulations of adsorption
\cite{FrenkelSmit}. The chemical potentials of the components can be specified in two ways. First, the pressure of the gas in the reservoir and its composition can
be specified. The Peng-Robinson equation of state is used to calculate the chemical potential or the fugacity of each component. This method was used to obtain
single component adsorption isotherms and $x$-$y$ diagrams for mixtures. The pressure varied in the range between 10$^4$ Pa and 10$^9$ Pa in these simulations.
Second, one can specify the fugacities of the components directly using the Lewis-Randall rule \cite{moranshapiro}. According to the Lewis-Randall rule, the
fugacity of each component equals the product of its mole fraction and the pure component fugacity at the same temperature and pressure of the mixture. This method
was used to obtain the adsorption isotherms for mixtures. This allowed us to perform the integration of \eqr{eq/FreeEnergy/06} along each chemical potential,
keeping the chemical potentials of the other components constant. The fugacity of each of the components varied in the range between 10$^4$ Pa and 10$^9$ Pa.

\section{Results of the simulations}\label{sec/Results}
\subsection{Single component adsorption isotherms}
\begin{figure}[h!]
\centering
\subfigure[ ] %
{\includegraphics[scale=\scaleprofile]{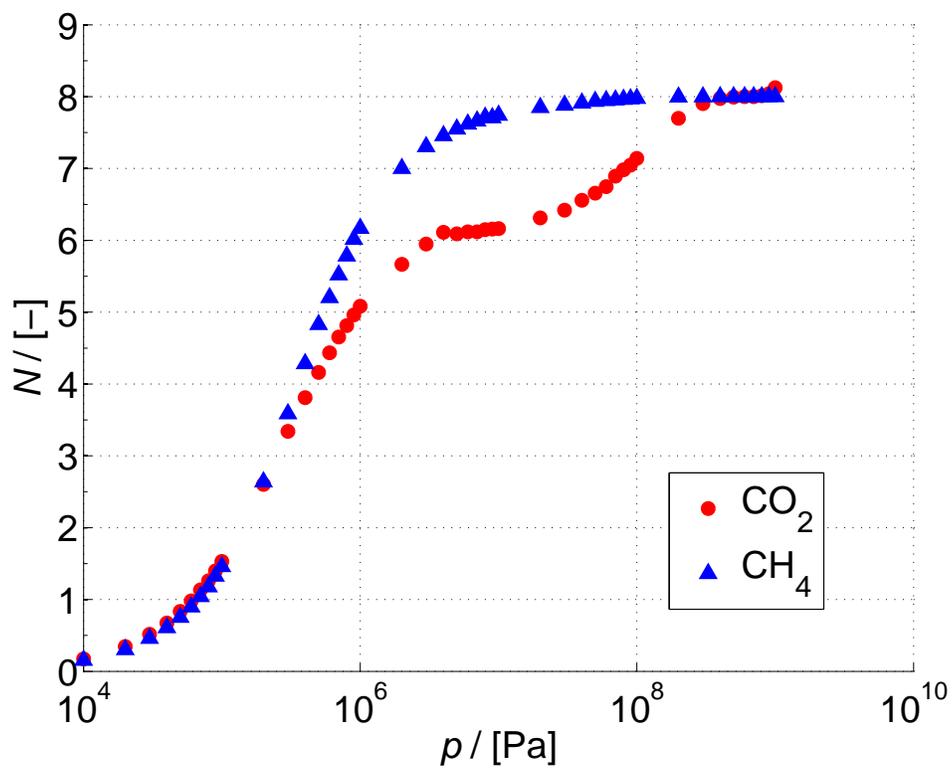}\label{fig/NP}} %
\subfigure[ ] %
{\includegraphics[scale=\scaleprofile]{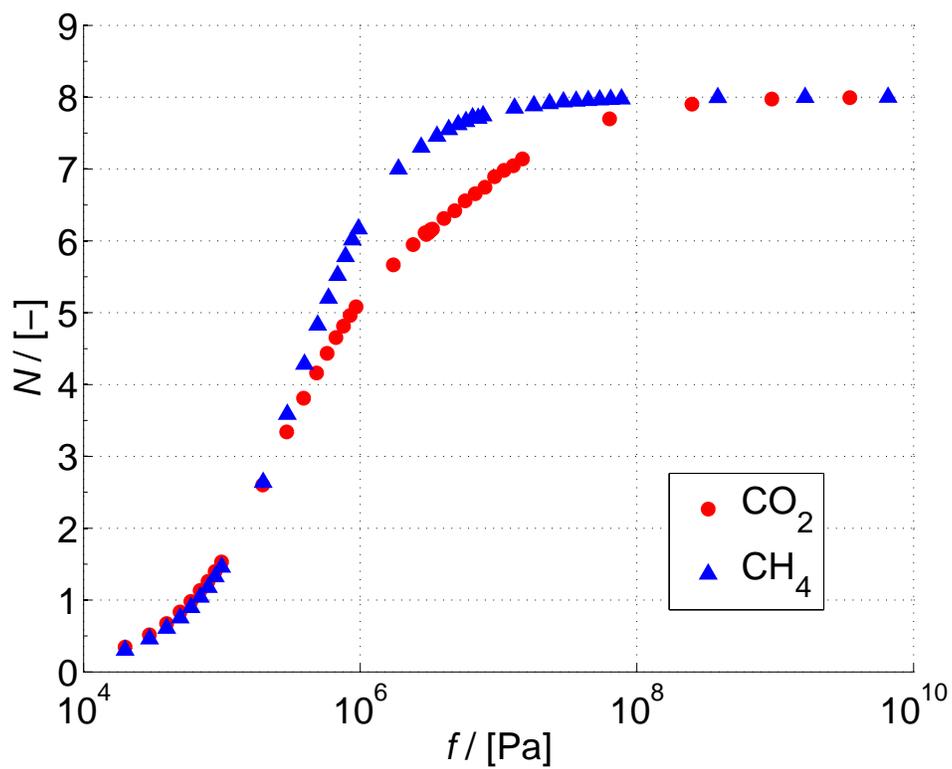}\label{fig/NF_eos}} %
\caption{Number of adsorbed molecules per unit cell of a sI hydrate as a function of the applied pressure (a) and fugacity (b) as computed by GCMC simulations at $T
= 278$ K.}\label{fig/loading}
\end{figure}

The single component adsorption isotherms for \co and \ch at 278 K are shown in \figr{fig/loading}. We see how the hydrate loading, $N$,  varies with applied
pressure $p$ on a sI unit cell with 8 cages. The pressure axis is given in a logarithmic scale. The error bar is less than the size of the symbol, for a hydrate
with immobile water molecules. The adsorption isotherms of \co and \ch for the case of mobile water molecules are also similar (not shown).

\begin{figure}[h!]
\centering
\subfigure[ ] %
{\includegraphics[scale=\scalevmd]{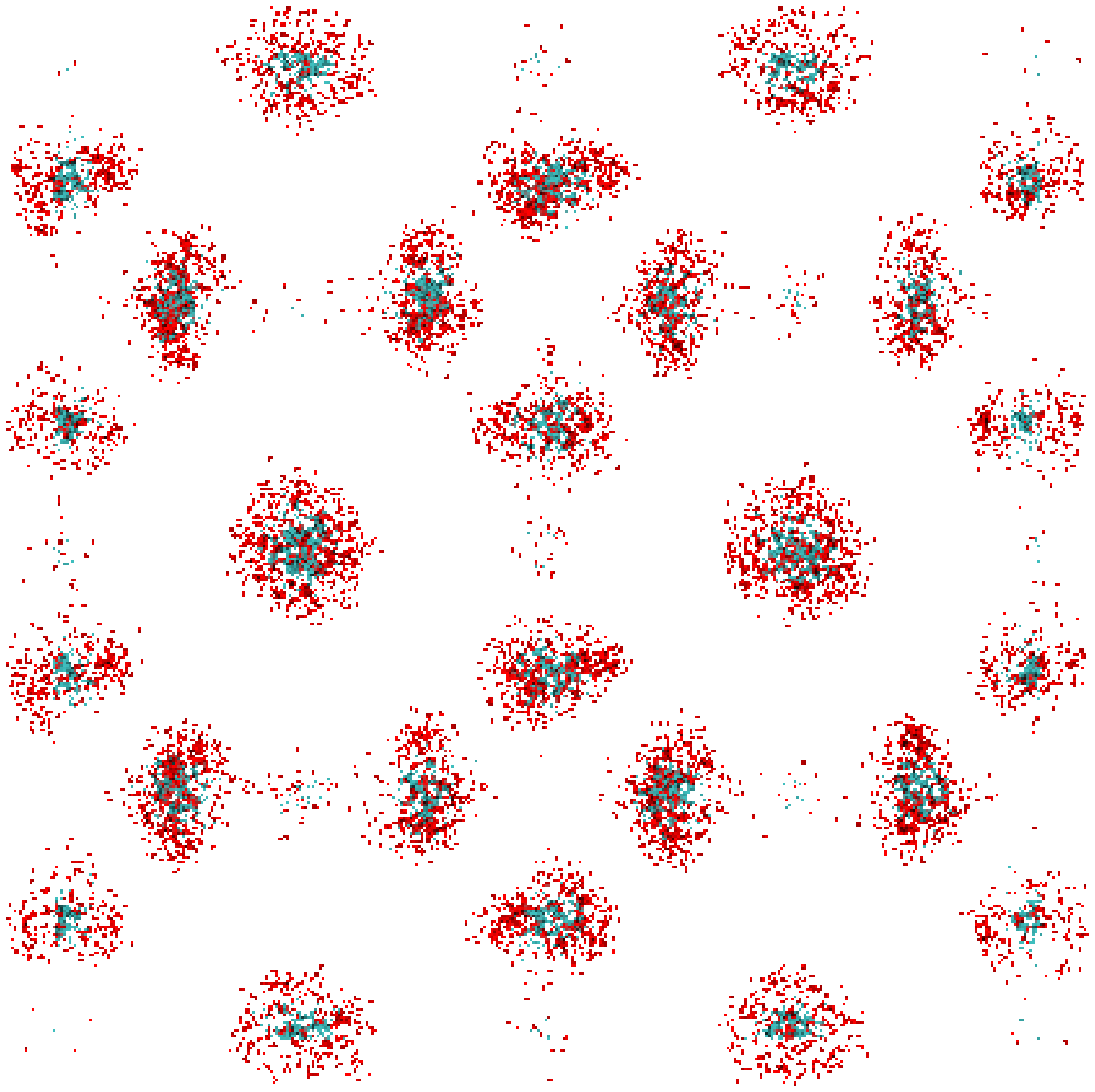}\label{fig/cages_CO2-11} } %
\subfigure[ ] %
{\includegraphics[scale=\scalevmd]{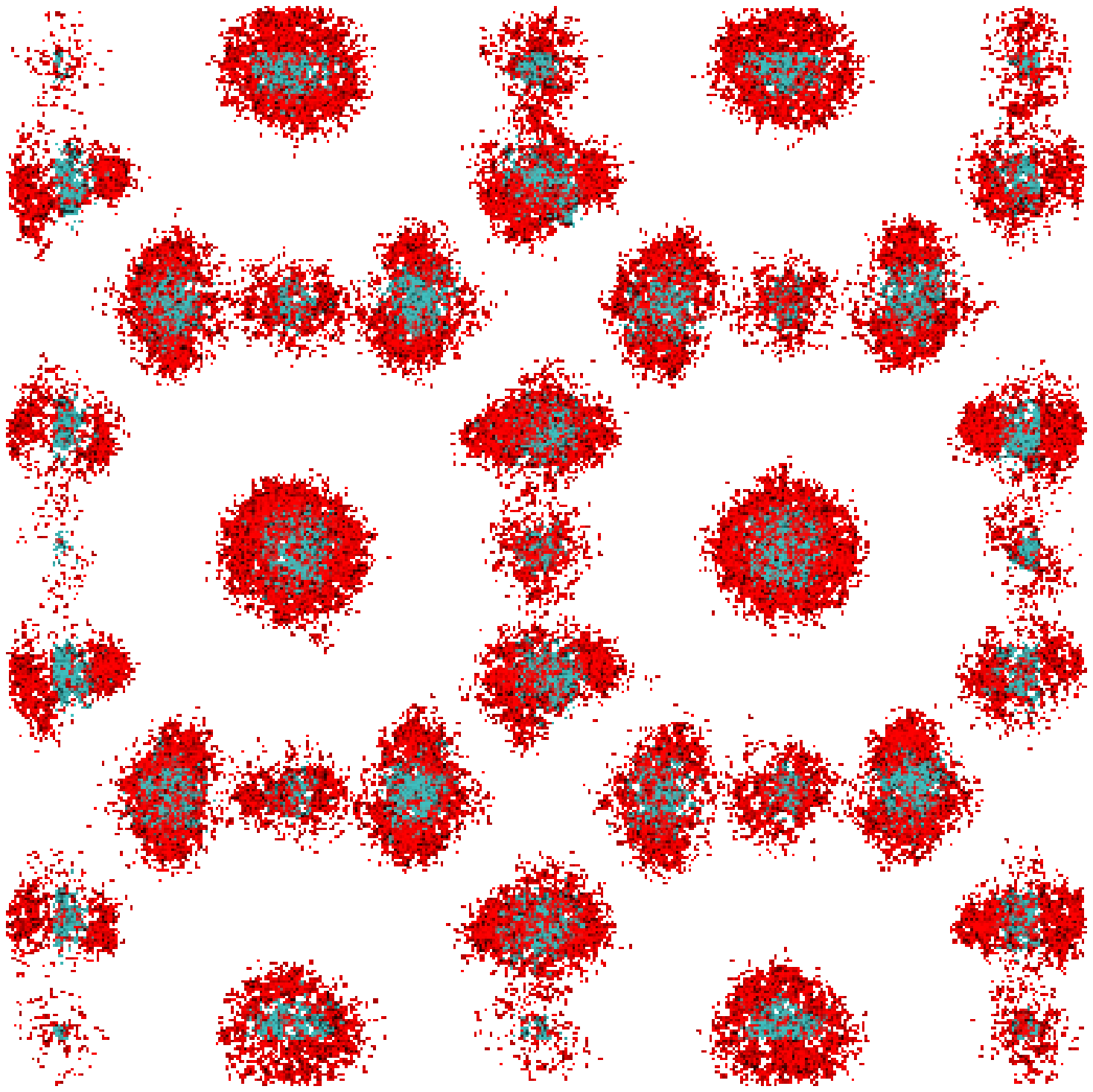}\label{fig/cages_CO2-52} } %
\subfigure[ ] %
{\includegraphics[scale=\scalevmd]{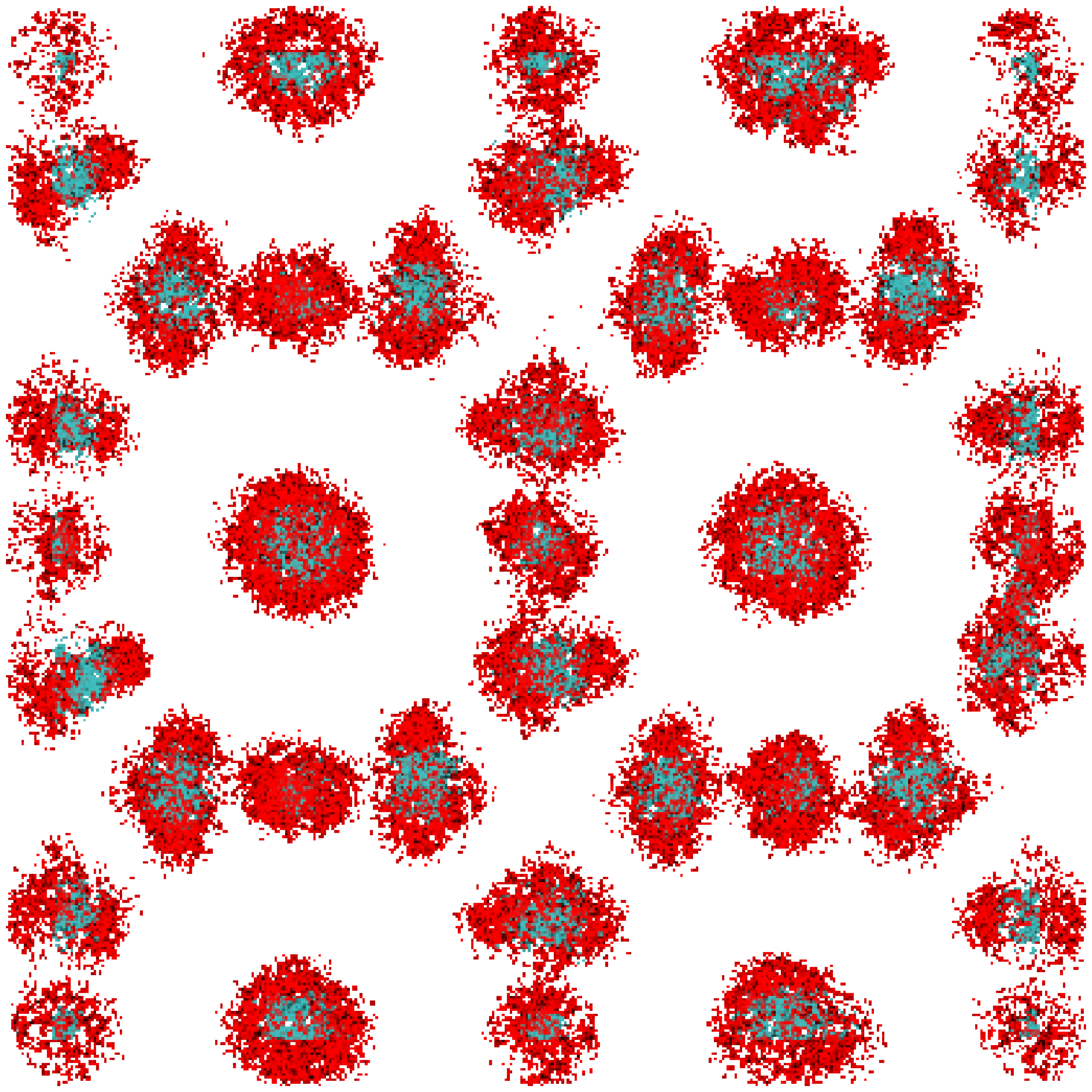}\label{fig/cages_CO2-66} } %
\caption{Distribution of \co molecules in the clathrate cages of a cell with 2x2x2 units of sI hydrate at various loading: (a) $N = 11$ molecules, (b) $N = 52$
molecules, (c) $N = 66$ molecules. Carbon atoms are colored blue, oxygen atoms are colored red, water molecules are not displayed for clarity.}\label{fig/cages_CO2}
\end{figure}

At first sight, the adsorption isotherms in \figr{fig/NP} may look like Langmuir adsorption isotherms, however the fitting of the one- or two- site profiles to such
isotherms is not satisfactory. Still, the \co isotherm reveals a two-site adsorption behavior. This can be explained by the structure of sI clathrate. A unit cell
of an sI hydrate has 6 large and 2 small cages. Being a relatively large molecule, \co tends to occupy large cages first. Small cages start to be filled only after
the large cages have been occupied. This can be also seen on \figr{fig/cages_CO2}. This figure shows the distribution of \co molecules between several cages,
obtained from a NVT simulation. The positions of these molecules from 500 snapshots were combined and drawn on a single figure. When the loading of \co is small ($N
= 11$), the loading of the large cages is much larger then the population of small cages. At intermediate loading ($N = 52$), which corresponds to the plateau in
the \co adsorption isotherm, some small cages start to be filled, while the other small cages remain almost unoccupied. Finally, at high loading ($N = 66$), all the
cages are occupied uniformly.

The cell with 2x2x2 units of sI clathrate has 64 cages. Having 66 molecules adsorbed into clathrate means that some cages contain 2 \co molecules. This happen at
large pressures only. Double occupancy of a cage can be explained by the linear form of \co molecule, which allows special orientation in large cages.

\begin{figure}[h!]
\centering
\subfigure[ ] %
{\includegraphics[scale=\scalevmd]{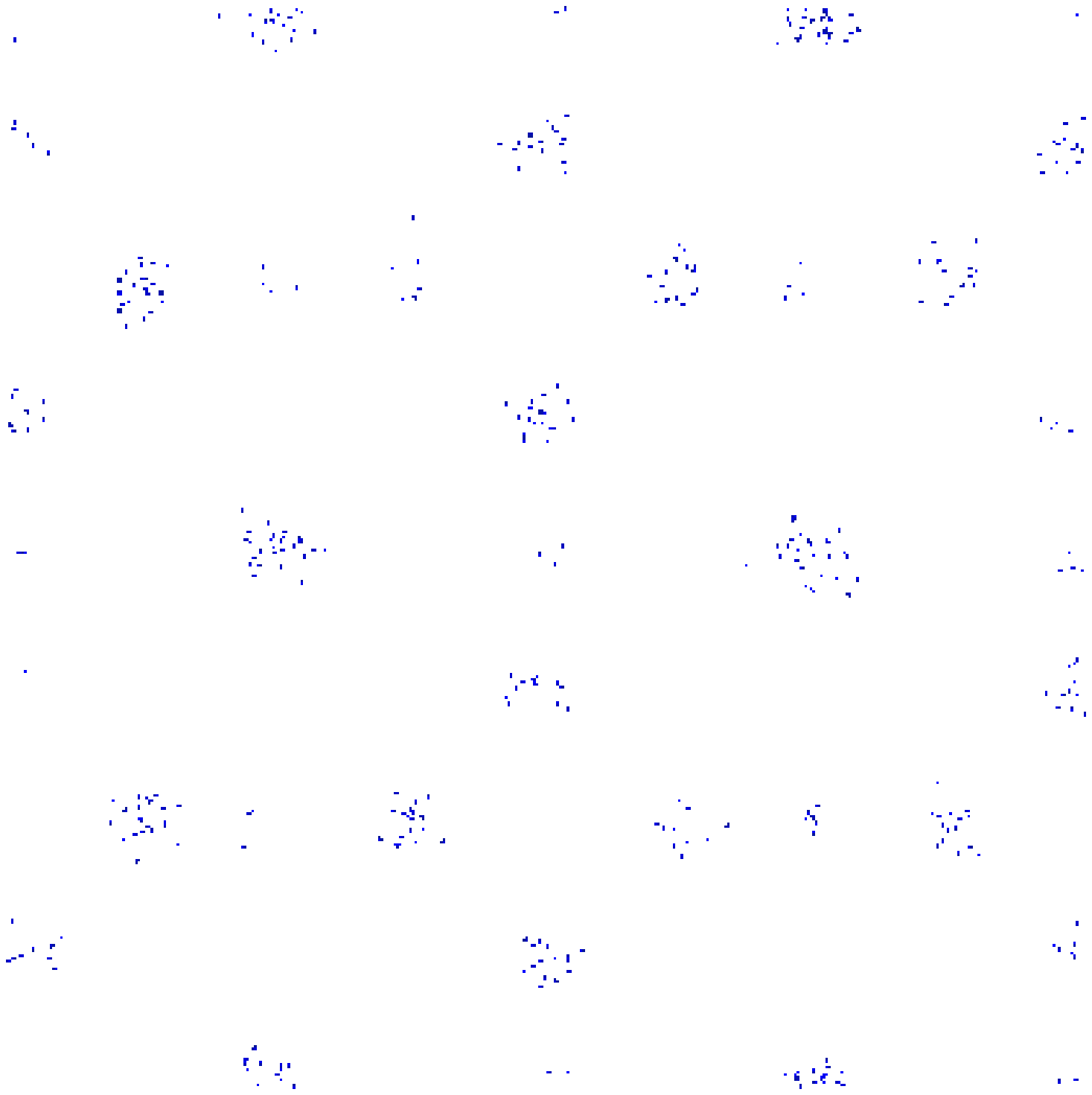}\label{fig/cages_CH41e+4} } %
\subfigure[ ] %
{\includegraphics[scale=\scalevmd]{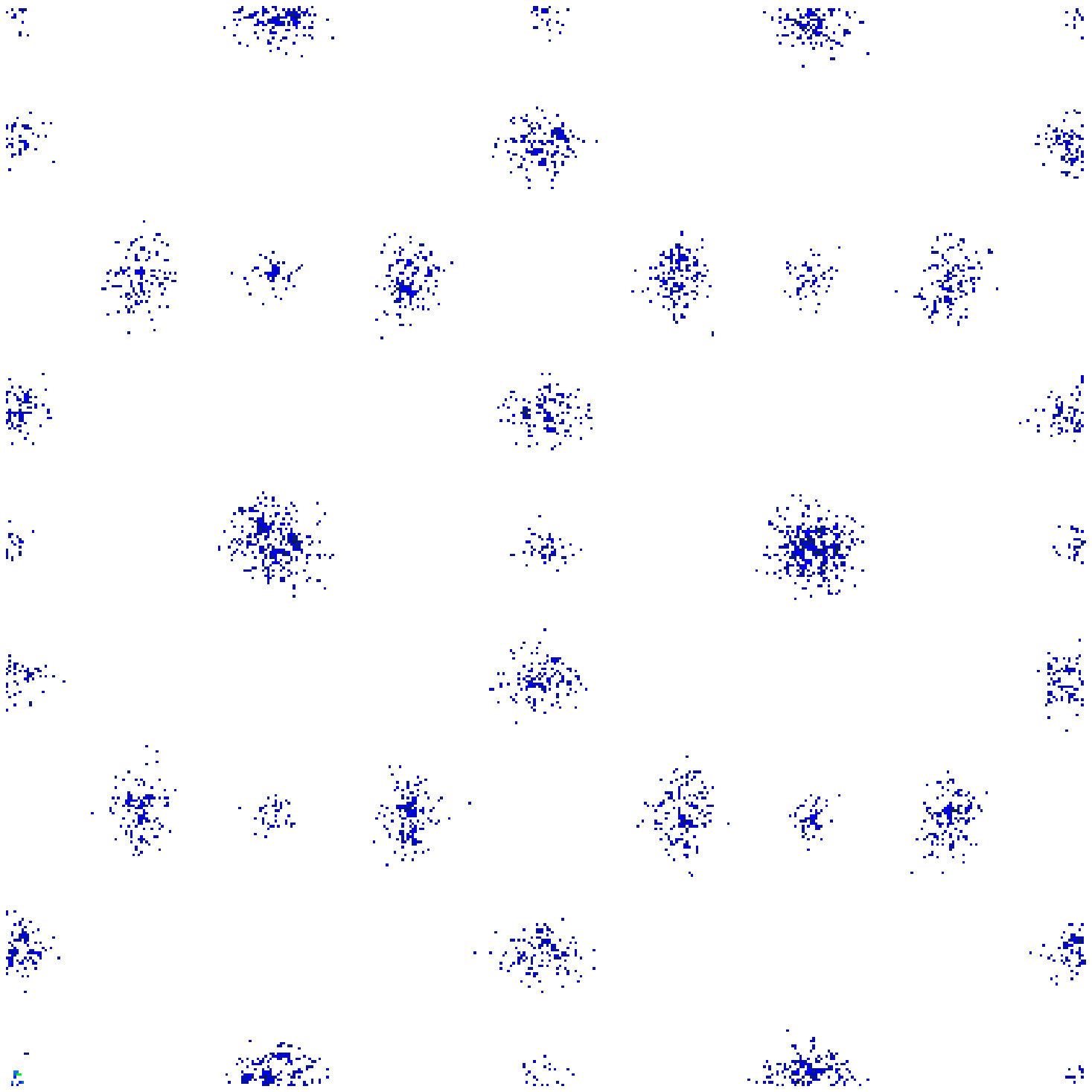}\label{fig/cages_CH41e+5} } %
\subfigure[ ] %
{\includegraphics[scale=\scalevmd]{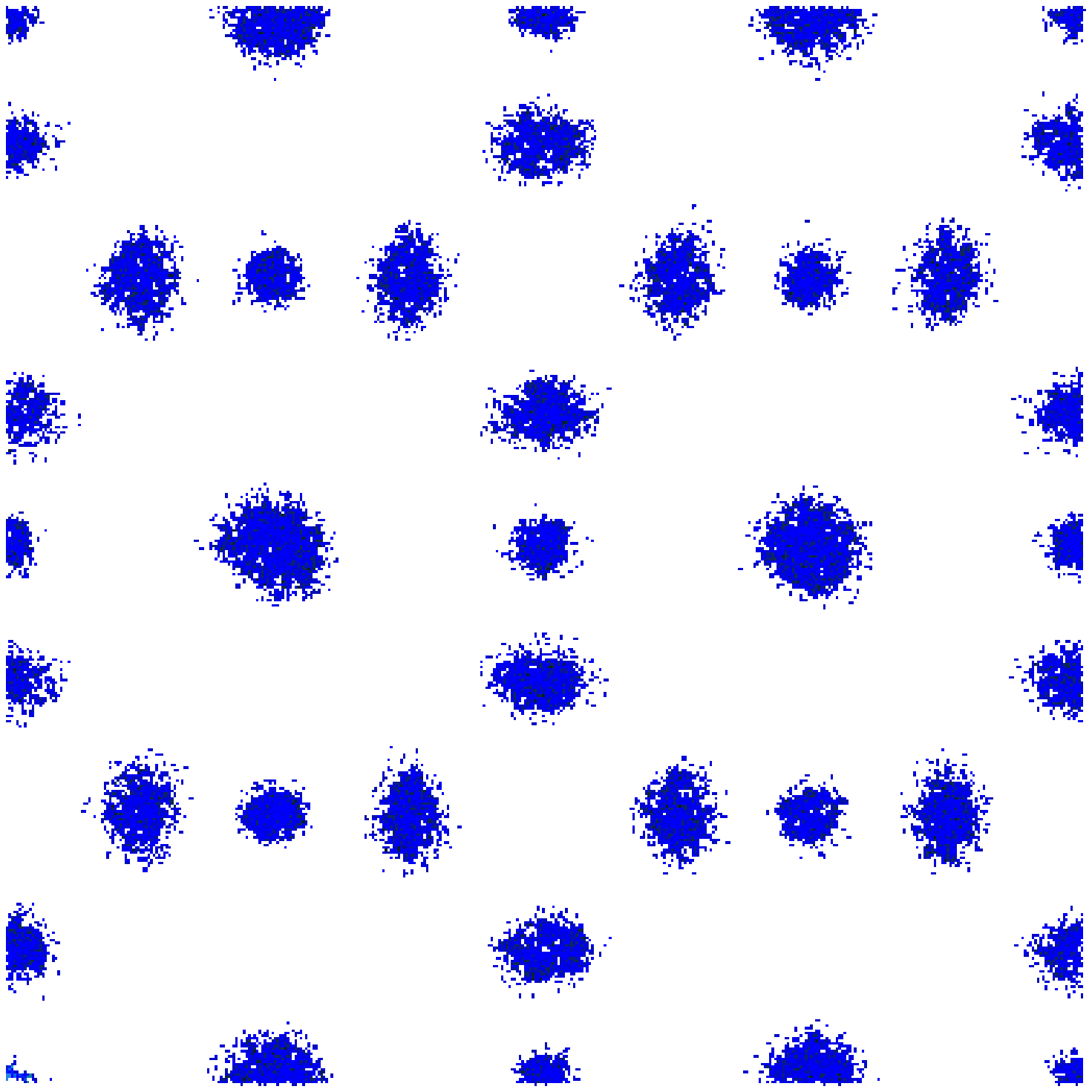}\label{fig/cages_CH41e+8} } %
\caption{Distribution of \ch molecules in the clathrate cages of a cell with 2x2x2 units of sI hydrate at various loading: (a) $p = 10^4 $ Pa, (b) $p = 10^5 $ Pa,
(c) $p = 10^8 $ Pa. \ch\, molecules are colored blue, water molecules are not displayed for clarity.}\label{fig/cages_CH4}
\end{figure}

The \ch isotherm, on the other hand, reveals a one-site adsorption behavior. The reason for this is that \ch molecules are rather small and have almost no
preference between large and small cages. However, the cage size is not completely irrelevant for the \ch distribution: otherwise the adsorption isotherm would be a
perfect single-site Langmuir curve. The typical distribution of \ch molecules is shown on \figr{fig/cages_CH4}. The snapshots were obtained in the same way as those
on \figr{fig/cages_CO2} from $\mu$VT simulations at different gas pressures. This figure shows that \ch fills the cages uniformly showing almost no preference to
large or small cages, in contrast to CO$_2$.

\begin{figure}[h!]
\centering
\subfigure[ ] %
{\includegraphics[scale=\scaleprofile]{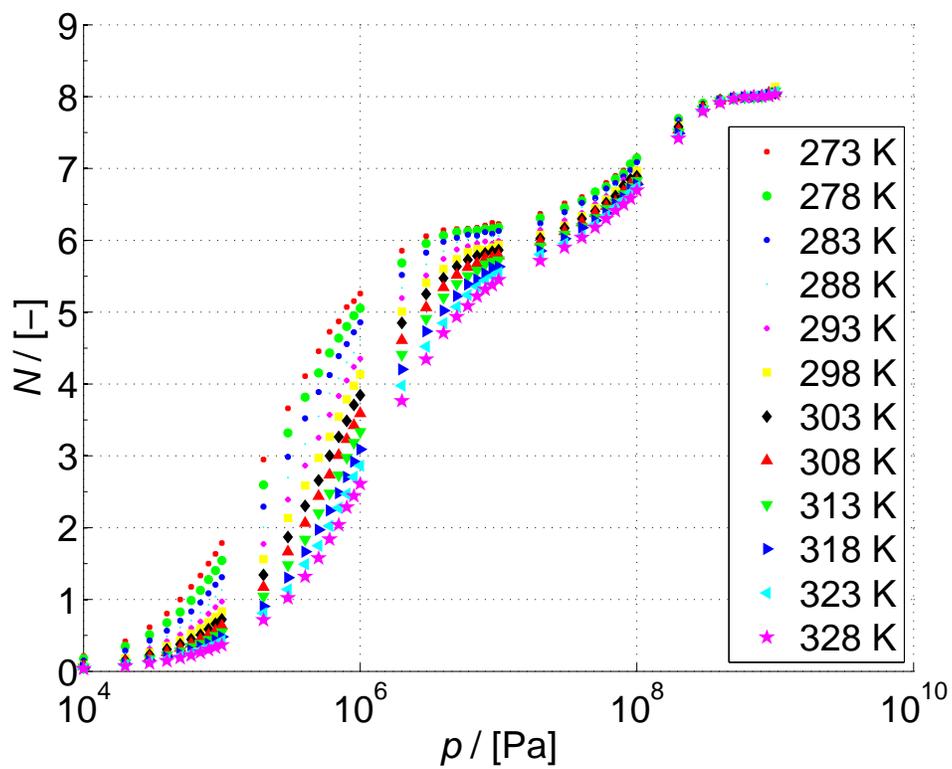}\label{fig/T_NP-CO2}} %
\subfigure[ ] %
{\includegraphics[scale=\scaleprofile]{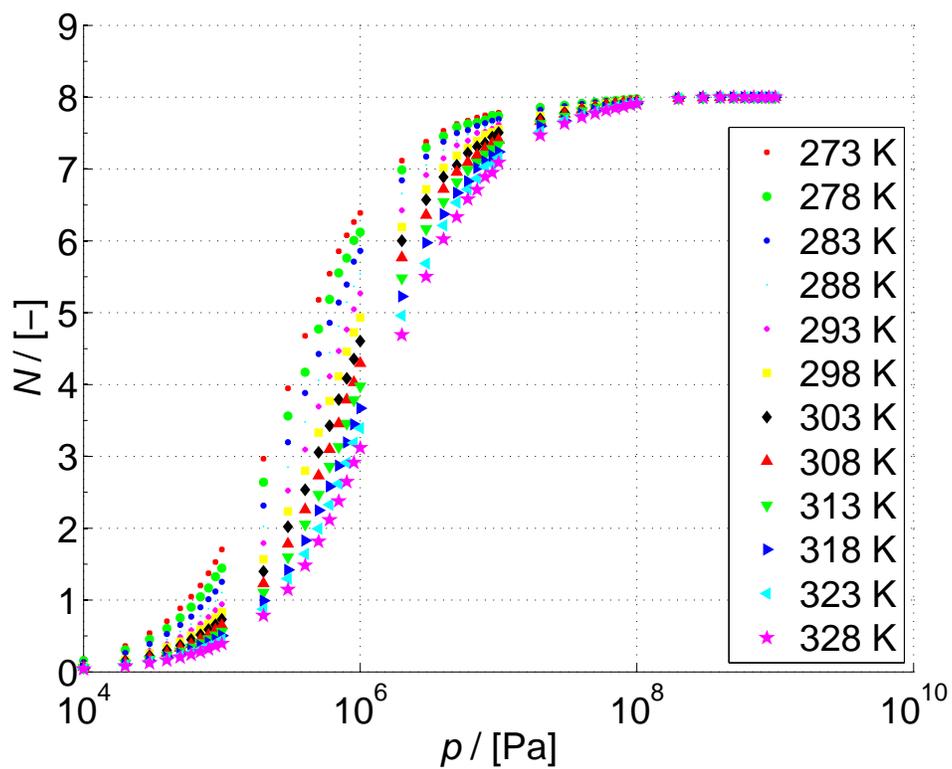}\label{fig/T_NP-CH4}} %
\caption{Number of adsorbed molecules per unit cell of a sI hydrate as a function of the applied pressure of (a) \co (b) \ch, as computed by GCMC simulations at
different temperatures.}\label{fig/T_NP}
\end{figure}

Furthermore, we computed the adsorption isotherms for different temperatures in the range between 273 K and 328 K. The results of simulations are presented on
\figr{fig/T_NP}. One can see, that increasing the temperature leads to the decrease in the amount of adsorbed molecules both of \co and CH$_4$. Nevertheless, the
hydrate becomes completely filled at pressures about $10^9$ Pa. It is interesting to analyze the distribution of the \co molecules in cages with respect to the
temperature. Recall that if the loading is less then 6, then mostly the large cages are being filled, while the higher loading correspond to filling also small
cages. We see from \figr{fig/T_NP-CO2} that increasing the temperature makes the isotherm to cross the line $N=6$ at higher pressure. Furthermore, the plateau at
$N=6$ becomes less pronounced when the temperature is increased. This shifts the type of adsorption isotherm from 2-site to 1-site, which means that the difference
between cages becomes less, when the temperature is increased. The \ch adsorption isotherm is qualitatively the same at different temperatures. Increasing the
temperature only increases the pressure at which the desired loading is achieved.

\subsection{Mixture adsorption isotherms}

\begin{figure}[h!]
\centering
\includegraphics[scale=0.6]{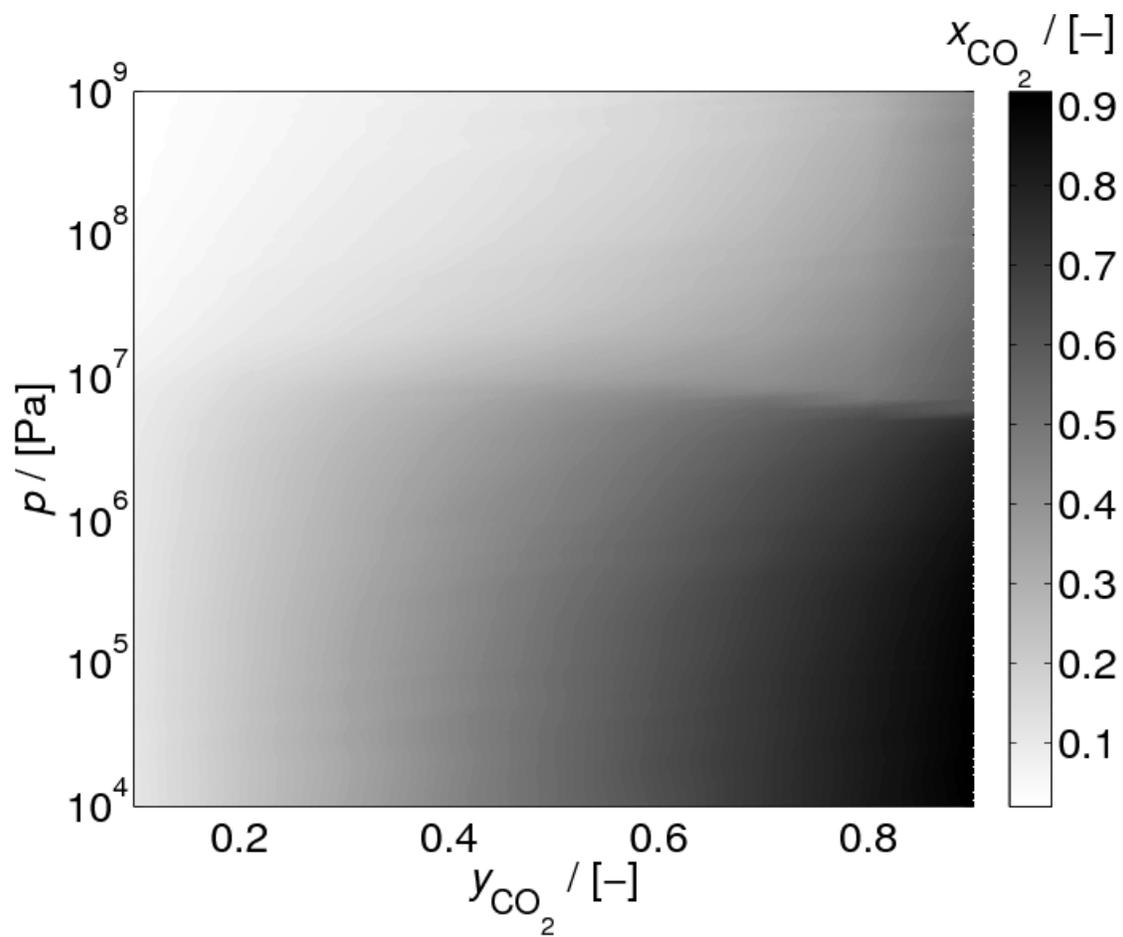}
\caption{Mole fraction of \co (color) in the \co+\ch gas mixture in the hydrate at 278 K computed by GCMC simulations as a function of the gas pressure and the mol
fraction of \co in the gas described by the Peng-Robinson equation of state.}\label{fig/XPY_CO2+CH4-01}
\end{figure}

\begin{figure}[h!]
\centering
{\includegraphics[scale=\scaleprofile]{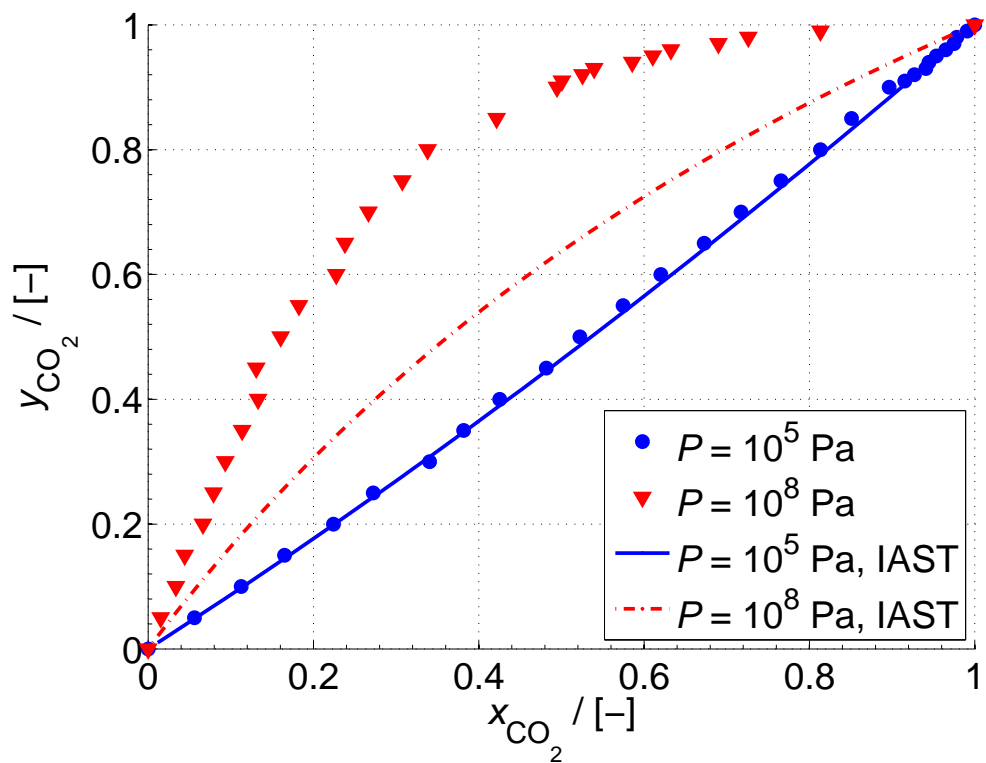}} %
\caption{Mole fraction of \co in the gas mixture, $y$, as a function of its mole fraction in the hydrate, $x$, at 278 K for various gas pressures, as computed by
GCMC simulations and compared to the predictions of ideal adsorption theory (IAST).}\label{fig/XY}
\end{figure}

Isotherms for hydrates loaded with mixtures of \co and \ch are shown in \figr{fig/XPY_CO2+CH4-01}. This figure shows the \co mole fraction in the hydrate as a
function of the gas pressure and the corresponding \co mole fraction in the gas phase. One can identify two regions in the figure, with the pressure above
approximately $10^7$ Pa and below $10^7$ Pa. As one can see from \figr{fig/NP} this is approximately the pressure where the adsorption isotherm of \co reaches the
plateau. In both regions the molar content of \co in the hydrate increases gradually when the content of \co in the gas phase increases. When the pressure is below
$10^7$ Pa the \co molecules fill the large cages while the \ch molecules fill both the small and the large cages. Thus, there is no preference between large and
small cages. Therefore, the mixture composition in the hydrate is approximately the same as in the gas phase over the whole range of compositions, see
\figr{fig/XY}. At approximately $10^7$ Pa the large cages are all filled and only small cages are available. \co molecules cannot compete with \ch molecules for
these cages, since the former ones are larger than the latter ones. It is mostly \ch molecules which occupy the small cages in the mixture, see \figr{fig/XY}.

\ref{fig/XY} compares the results of the mixture loading with ideal adsorption solution theory (IAST) \cite{IAST}. The concept of ideal solution for the adsorbed
components is analogous to the Raoult's law for vapor-liquid equilibrium. It holds well for mixtures of similar components which are adsorbed on similar sites. For
segregated systems (when one component adsorbs on one type of sites while the other component adsorbs on the other adsorption sites) it is known that it does not
work \cite{Krishna2001, Murthi2004}. As one can see from \figr{fig/XY}, the mixture can be considered ideal at small pressures, which are below the pressure when
\co loading reaches the plateau. At these pressures \co and \ch molecules occupy the large cages equally likely. Even though \ch and \co molecules are different,
the large cages are so large, that they show no preference to the components they would adsorb. At pressures larger than $10^7$ Pa we observe preferential
adsorption for \ch molecules, so the mixture cannot be considered ideal. At these pressures the small cages are being filled, so it is natural that they prefer
small \ch molecules rater than large \co molecules. It becomes clear that at high pressures IAST does not hold.

\begin{figure}[h]
\centering
\subfigure[] %
{\includegraphics[scale=0.6]{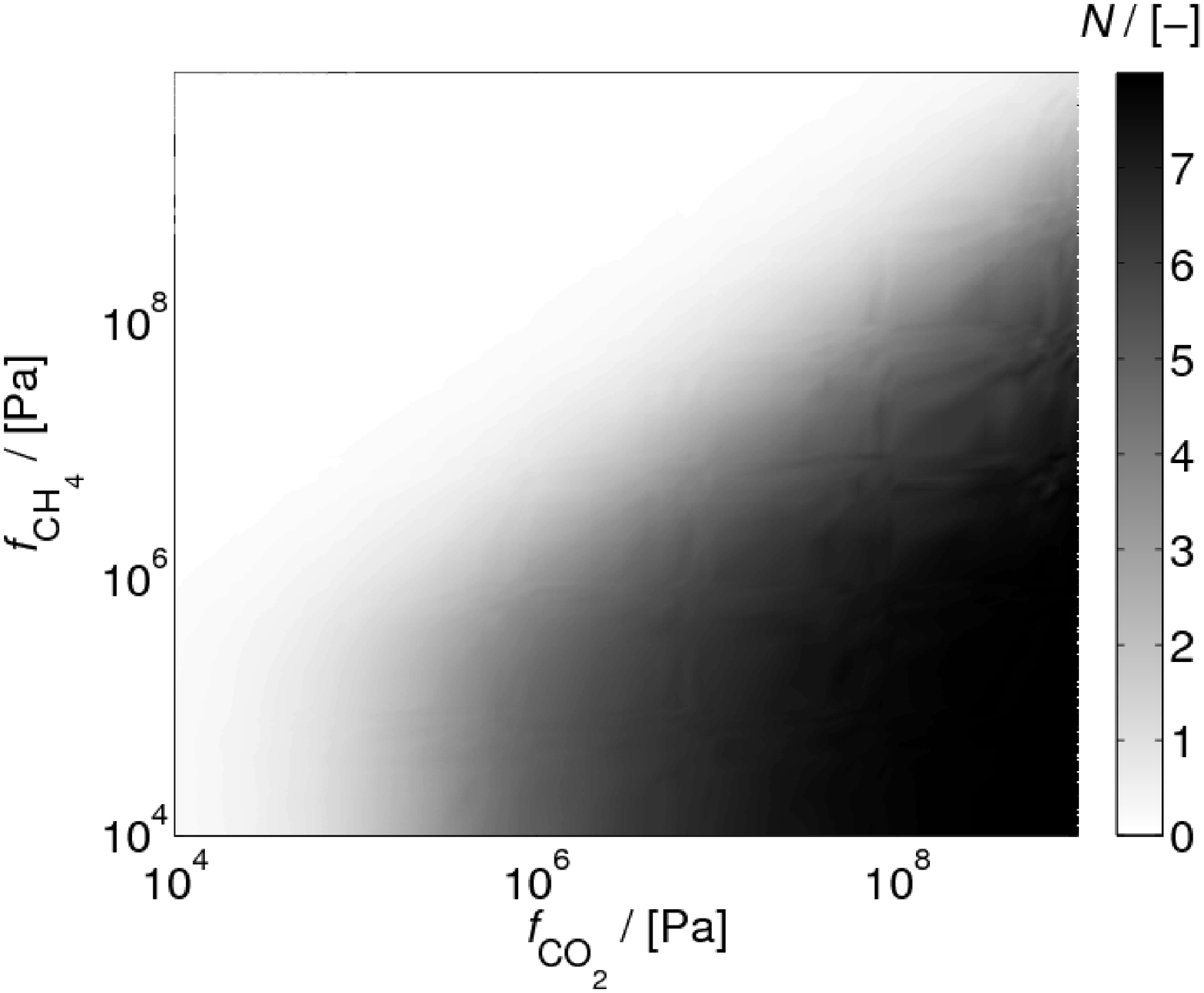}\label{fig/NFF_CO2} } %
\subfigure[] %
{\includegraphics[scale=0.6]{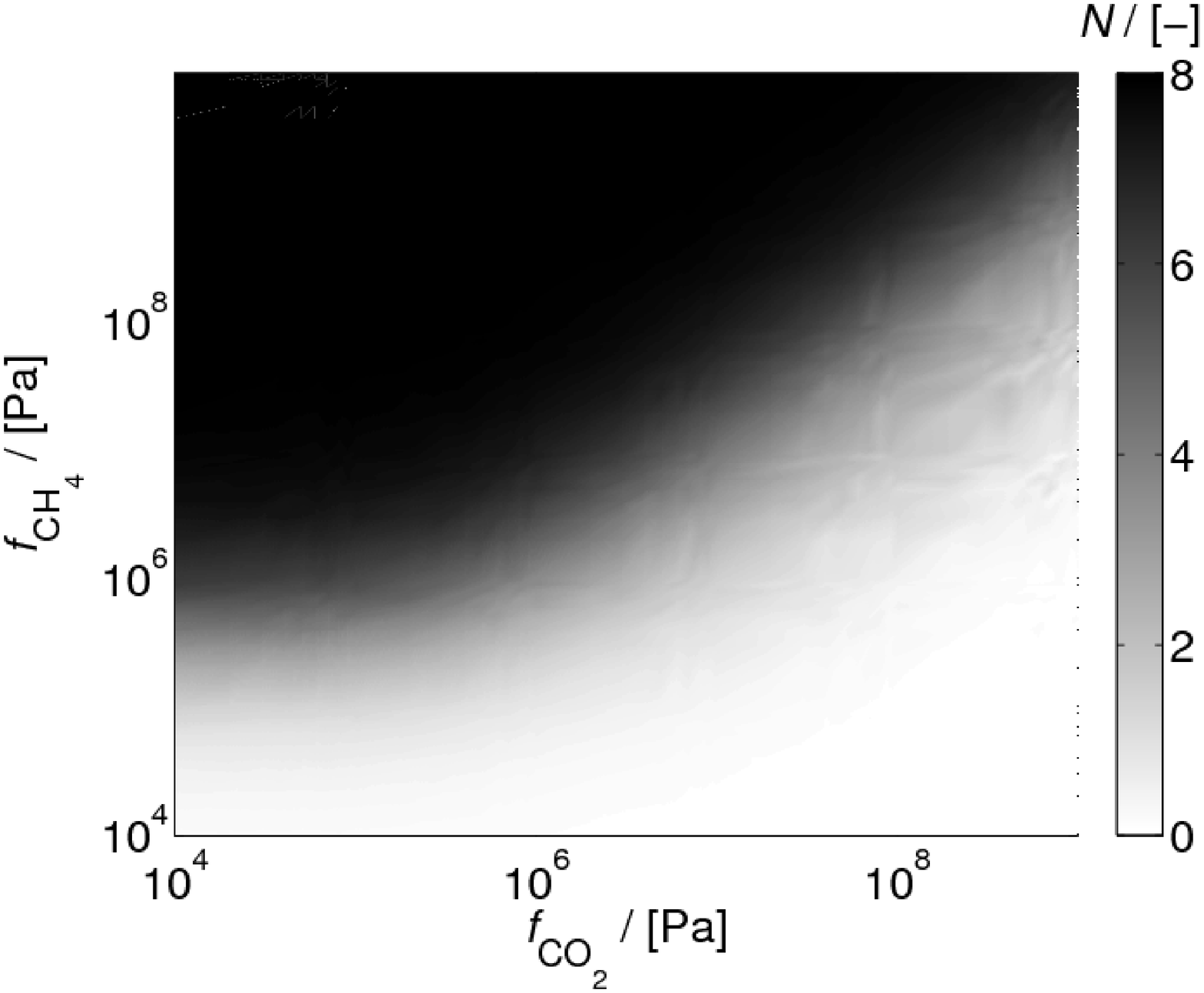} \label{fig/NFF_CH4} } %
\caption{Loading per unit cell (color) of each of the component in the \co+\ch gas mixture in the hydrate at 278 K as a function of the fugacity of each of the
component: (a) loading of \co, (b) loading of \ch.}\label{fig/NFF}
\end{figure}

We next consider the dependence of loading on the fugacity of each of the component, which is shown on \figr{fig/NFF}. We see that the loading of each of the
component increases proportionally to its fugacity. Furthermore, the increase of the fugacity of one of the components complicates the adsorption of the other
component and vice versa. This explains the dark triangular region in the bottom-right and the top-left corners of the figure for \co and \ch respectively.
Furthermore, one can see the gradual filling of the hydrate with the increase of the total fugacity $f_{\coeq}+f_{\cheq}$ which is related to the gas pressure. This
behavior is consistent with the adsorption isotherms on \figr{fig/NF_eos}.

\subsection{The Helmholtz energy of the pure gases and gas mixtures in hydrates}

\begin{figure}[h]
\centering
\includegraphics[scale=\scaleprofile]{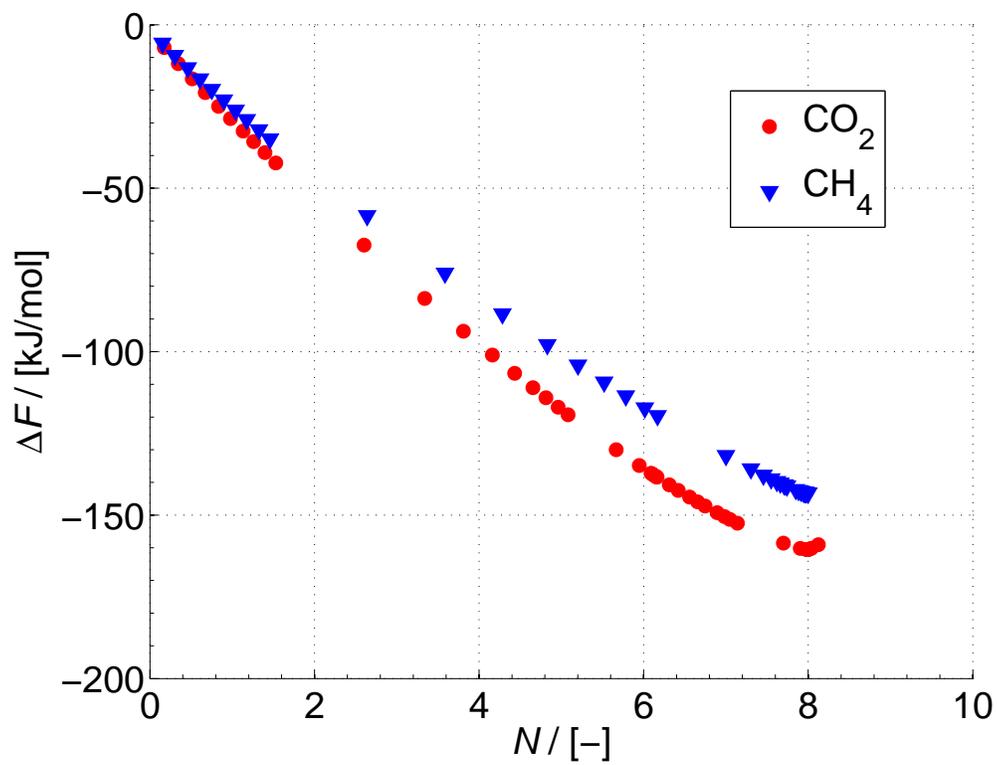}
\caption{Helmholtz energy of a single component hydrate at 278 K as a function of the hydrate loading per unit cell.}\label{fig/FN}
\end{figure}

We first report the Helmholtz energy for the single component hydrates. \figr{fig/FN} shows this Helmholtz energy for \co and \ch hydrates. One can see, that the
more molecules are adsorbed, the lower is the Helmholtz energy per mole. This means that the hydrate is becoming relatively more stable as the number of molecules
increases. This is well known \cite{SloanHydrates}, because empty cavities are not stable and guest molecules are required to stabilize them. Furthermore, one can
see that the Helmholtz energy for the \co hydrate is lower than that of the \ch hydrate over the whole range of loadings. This indicates that the \co hydrate is
more stable than the \ch hydrate.

\begin{figure}[h]
\centering
\subfigure[] %
{\includegraphics[scale=\scaleprofile]{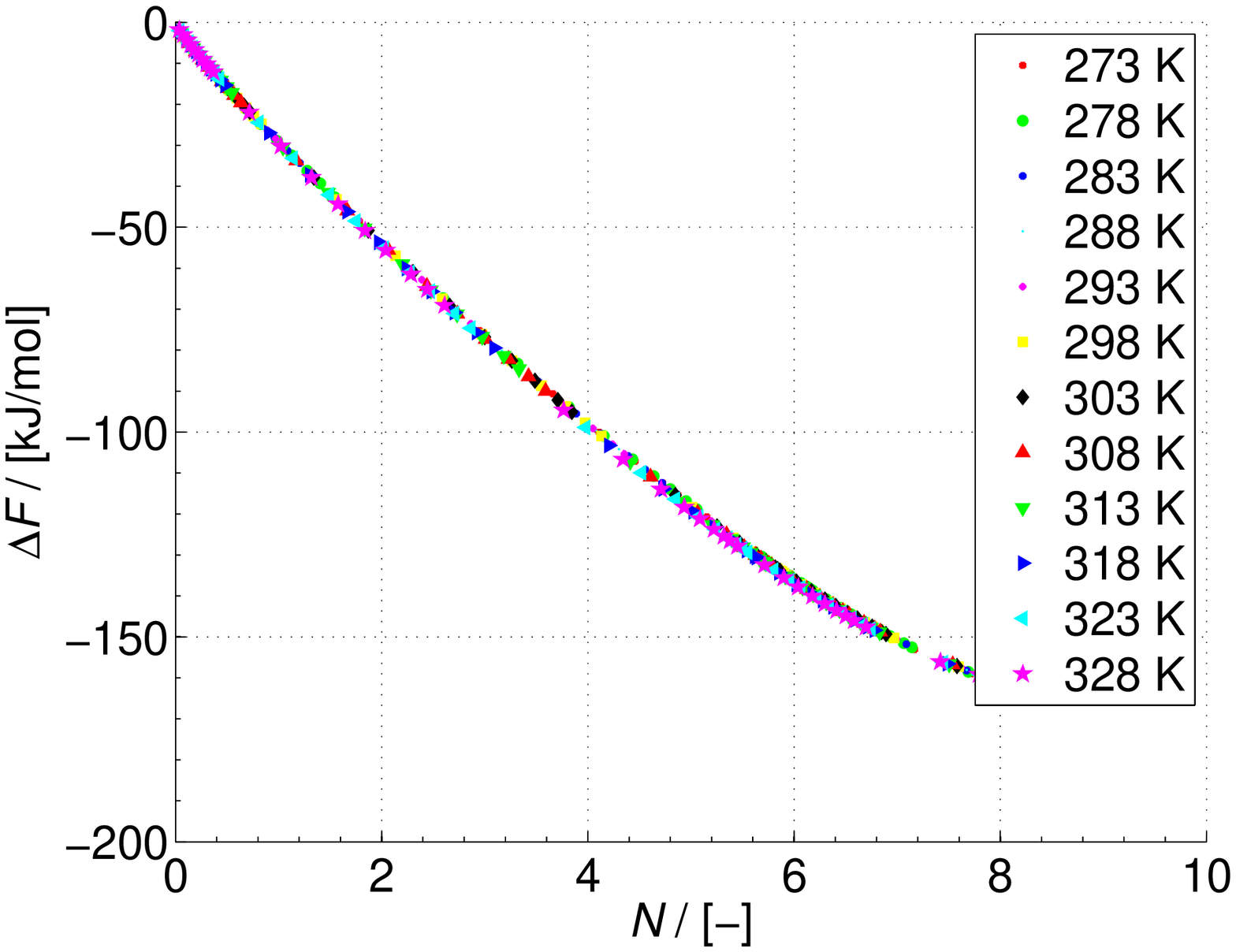}\label{fig/T_FN-CO2} } %
\subfigure[] %
{\includegraphics[scale=\scaleprofile]{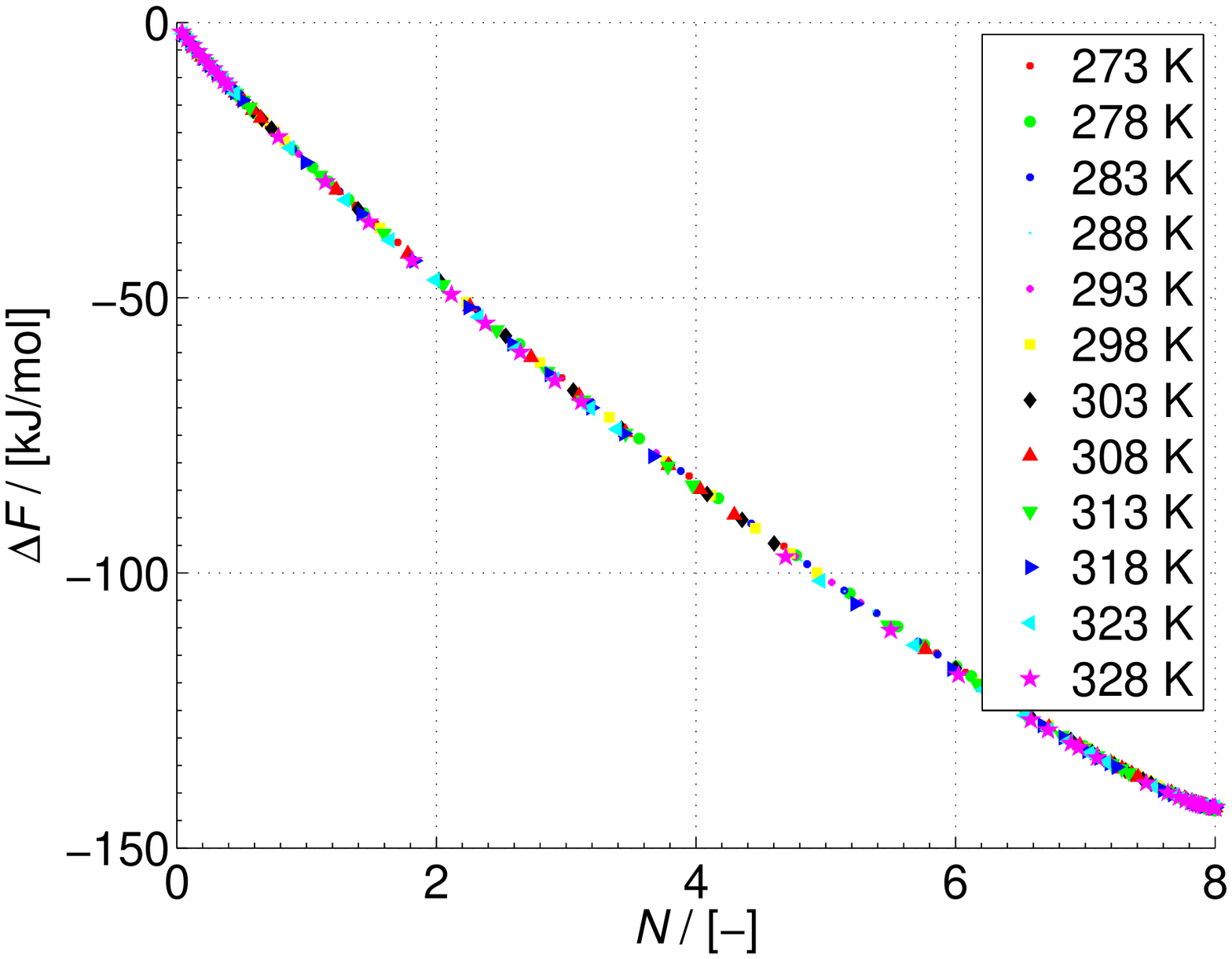} \label{fig/T_FN-CH4} } %
\caption{Helmholtz energy of a single component hydrate at different temperatures as a function of the hydrate loading per unit cell for (a) \co (b) \ch.}
\label{fig/T_FN}
\end{figure}

Furthermore, we consider the temperature dependence of the Helmholtz energy of a single component hydrate. The computed data in the range of temperatures between
273 K and 328 K are presented on \figr{fig/T_FN}. One can see, that variation of temperature does not affect the Helmholtz energy at all. This means, that the
change of the entropy with respect to the change in loading is zero, $\Delta S = \left(\partial \Delta F/ \partial T\right)_{V, N} = 0$. This indicates that the
adsorbed molecules in different cages do not interact.

\begin{figure}[h]
\centering
\includegraphics[scale=0.6]{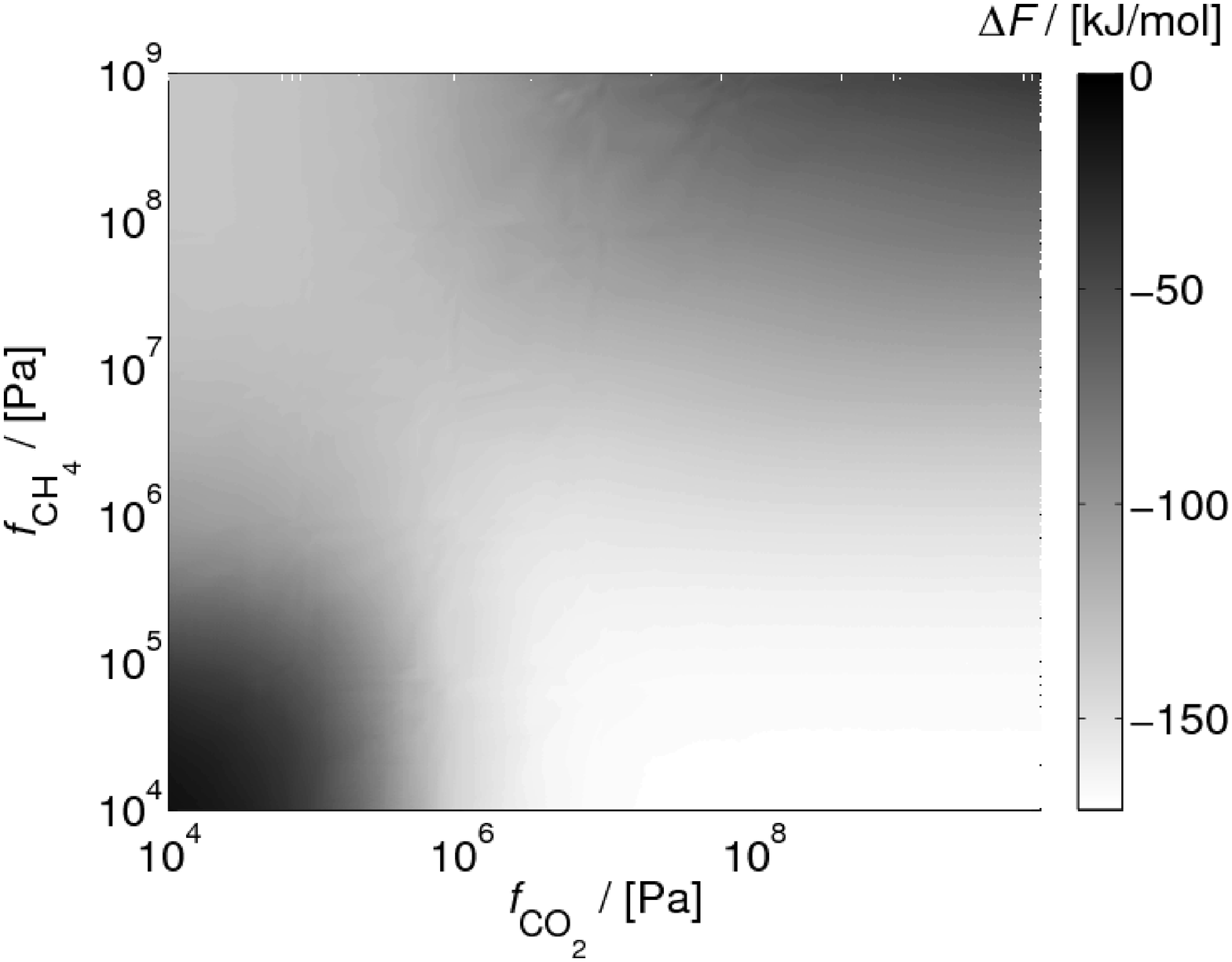}
\caption{Helmholtz energy difference (color) as computed by \eqr{eq/FreeEnergy/06} of the \co+\ch gas mixture in the hydrate at 278 K as a function of the fugacity
of each of the component.}\label{fig/FFF_CO2+CH4}
\end{figure}

We next consider the Helmholtz energy of the gas mixtures in hydrate. \figr{fig/FFF_CO2+CH4} shows how the Helmholtz energy depends on the fugacities of each of the
components. We see that there is a certain region where this energy is lowest, so the hydrate with gas mixture would be more stable at these composition. At low
fugacities there are not that many molecules adsorbed, which means that many cages are empty. This makes the gas hydrate structure unstable, which explains the dark
region on the figure in the bottom-left corner. If both fugacities are high, both \co and \ch molecules prefer to occupy the cages. Competition for an available
cage makes the hydrate structure less stable, as well. This is indicated by the dark region on the figure in the top-right corner. Finally, if only one of the two
fugacities is large, while the other is low, only the high fugacity molecules tend to fill the cages and the hydrate structure becomes relatively more stable.

These considerations of stability are not conclusive, however. As explained in the introduction a more complete set of thermodynamic data is needed to evaluate the
possibility of hydrate formation. Namely, the values of the Helmholtz energy of different phases which can be formed of CH$_4$, \co and water are required. We shall
return to this point in our next work.

\begin{figure}[h]
\centering
\subfigure[] %
{\includegraphics[scale=0.5]{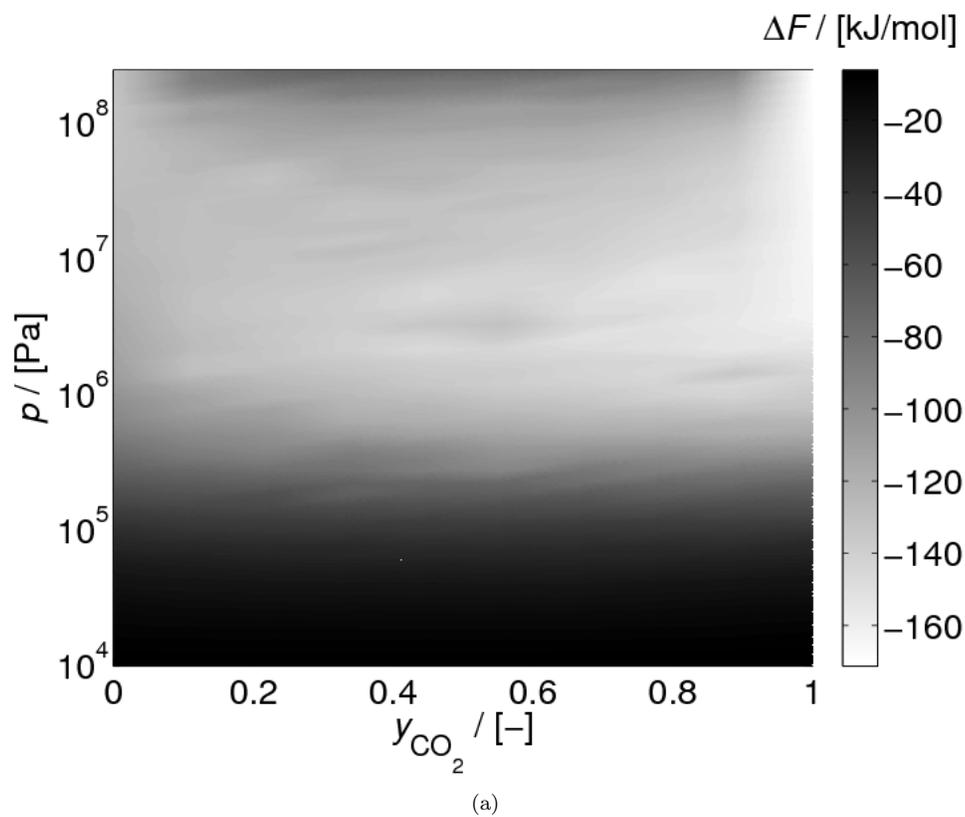}\label{fig/FPY_CO2+CH4}}
\subfigure[] %
{\includegraphics[scale=0.5]{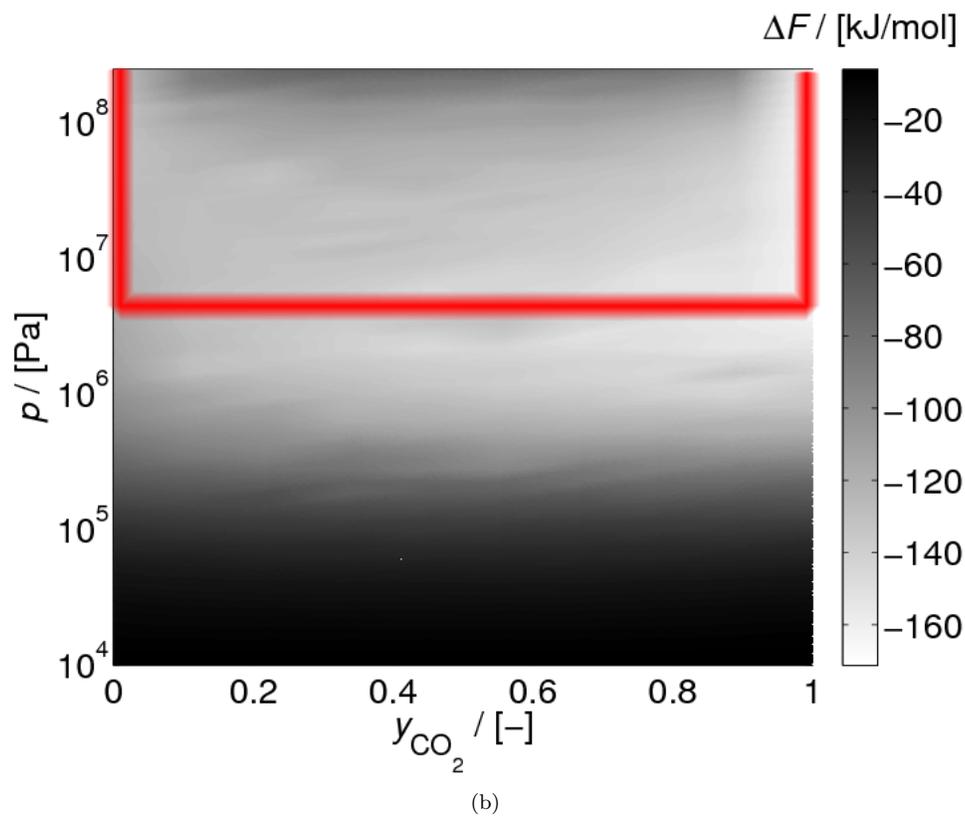}\label{fig/FPY_CO2+CH4_p}}%
\caption{(a) Helmholtz energy difference (color) as computed by \eqr{eq/FreeEnergy/06} of the \co+\ch gas mixture in the hydrate at 278 K as a function of the gas
pressure and mol fraction of \co in the gas phase. (b) The path on the diagram to convert a pure \ch hydrate to a pure \co hydrate.}
\end{figure}

It is still interesting to plot \figr{fig/FFF_CO2+CH4} in a $p$-$y$ scale, where $p$ is the gas pressure, which is in equilibrium with the guest molecules in the
hydrate, and $y$ is the mole fraction of \co in the gas phase. The distribution of the Helmholtz energy of the hydrate with gas mixture for this case is shown in
\figr{fig/FPY_CO2+CH4}. The pressure and the mole fraction were calculated from the fugacities of the components using the Lewis and Randall rule
\cite{moranshapiro}. We see that the hydrate is more stable at high pressures (i.e. at high loadings) if pure gas is being adsorbed. In contrast, if we have a
mixture, the highest stability is achieved at moderate pressures.

\ref{fig/FPY_CO2+CH4} shows that there is a region in the pressure-composition diagram, which connects a relatively stable methane hydrate with a relatively stable
carbon dioxide hydrate. While at low and high pressures the hydrate with gas mixture is unstable, it is almost as stable as either single component hydrate at
moderate pressures. This is very promising, since one of the goals of our research is to find a path on a phase diagram to convert a pure \ch hydrate into a pure
\co hydrate without destroying its structure. In a pursuit of this path it will therefore be interesting to focus on the lightest region on the diagram. On
\figr{fig/FPY_CO2+CH4_p} this path is drawn schematically. Note that following this path implies that one does not meet any energy barrier: the value of $\Delta F$
along it is almost everywhere constant. In contrast, going from pure \ch hydrate to pure \co hydrate at constant pressure around $10^{8}-10^{9}$ Pa requires
crossing an energy barrier of 50-100 kJ per mol of unit cell.

We see that this path goes through a region where a mixture of \ch and \co is adsorbed into a hydrate. Namely, one has first to decrease the pressure of pure \ch
hydrate down to approximately $10^7$ Pa. At this pressure there is enough space in the large cages for \co molecules to go into them. Next, one has to increase the
content of \co in the surrounding fluid. This would lead to the filling of the large cages with \co molecules while the \ch molecules will be expelled out of the
hydrate. When there is no \ch in the mixture, the small cages are empty at the pressure $10^7$ Pa, and this is the time to start filling them in with the \co
molecules. This is the last part of our path which ends up with a pure \co hydrate.

\subsection{Heats of adsorption}

\begin{figure}[h]
\centering
\includegraphics[scale=\scaleprofile]{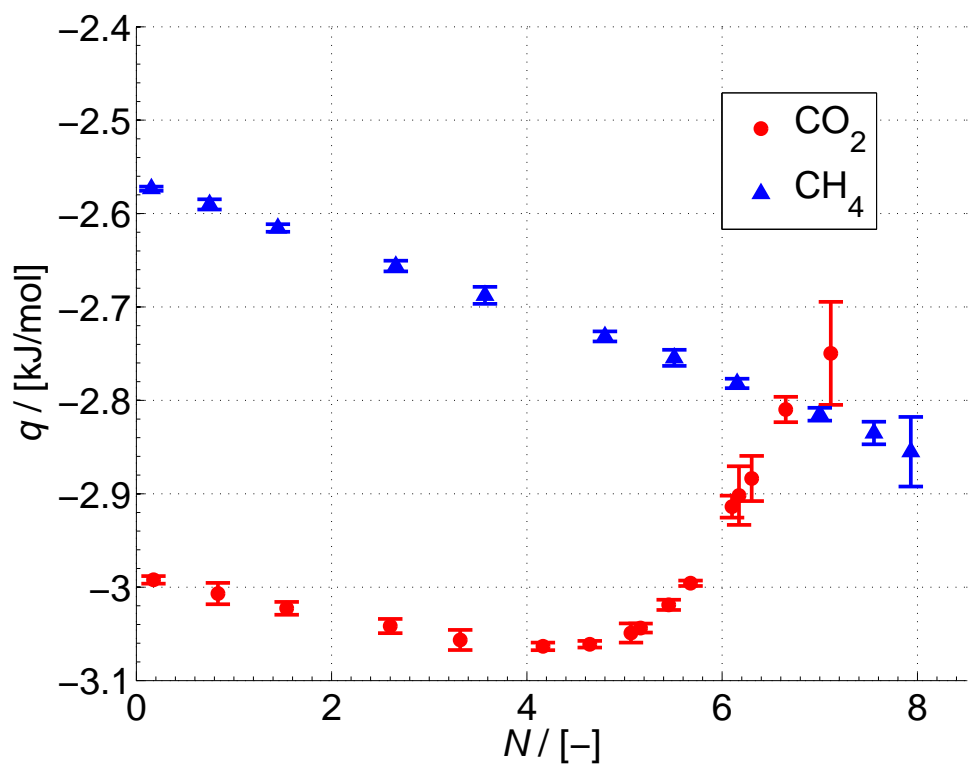}
\caption{Partial molar heat of adsorption of a single component hydrate as computed by GCMC simulations at 278 K as a function of the loading of a single-component
hydrate.}\label{fig/HN}
\end{figure}

\ref{fig/HN} shows the partial molar heats of adsorption, \eqr{eq/HeatOfAdsorption/06}, of the single component hydrates obtained from GCMC simulations. At high
loading ($N \approx 7$) the calculated value of the partial molar heat of adsorption starts to fluctuate. This is expected for GCMC simulations, since in the dense
phase which is realized at high loadings, the fraction of accepted trial MC moves is quite low. This means that that we shall trust the results at small and
moderate loadings only (not more then 6-7 adsorbed molecules per unit cell). Both sets of data agree with the findings above. 

We see in \figr{fig/HN} that the partial molar heat of adsorption of \ch decreases almost linearly with the loading over the whole range of loadings. We also know
that the change of the entropy $\Delta S$ with respect to the change in loading (see \eqr{eq/Intro/01}) is zero. The single-site adsorption isotherm shown on
\figr{fig/NP} for methane can thus be explained by \figr{fig/HN} alone. In contrast, for carbon dioxide one can distinguish the two regions for the partial molar
heat of adsorption of \co. At small loadings the partial molar heat of adsorption decreases, similar to that for CH$_4$. At this moment the large cages of the
hydrate are being filled in. At moderate loadings the partial molar heat of adsorption has a tendency to increase, however. This corresponds to the filling of the
small cages of the hydrate. As the \co molecules prefer not to be in the small cages, more and more energy is required to get them into hydrate at high pressures.
Knowing that $\Delta S = 0$ also here, the shape of the curve in \figr{fig/HN} can explain the adsorption isotherm for \co shown on \figr{fig/NP}. Indeed, the
profiles on \figr{fig/HN} at small loadings are not exactly linear, which implies that the adsorption isotherm is not Langmuir-like.

The profile for \co in \figr{fig/HN} changes direction not at $N=6$ which corresponds to the complete filling of the large cages. The minimum is when only 4-5 cages
are filled in. This may be explained by a strong interactions of large asymmetric \co molecule with water cages \cite{Kvamme2007a}.

The partial molar heat of adsorption of \co is lower than the one of \ch over the whole range of loadings which can be trusted. This supports the preferential
filling of the large cages by \co molecules.

Furthermore, we consider the temperature dependence of the partial molar heat of adsorption of a single component hydrate. The results of simulations in the range
of temperatures between 273 K and 328 K are presented on \figr{fig/T_FN}. Within the error of simulation one can consider the partial molar heat of adsorption to be
temperature independent, like the Helmholtz energy, computed in the previous section.

\section{Conclusions}\label{sec/Conclusions}

In this paper we studied the thermodynamic properties of an sI hydrate filled with CO$_2$, \ch and the mixture of those. By performing grand-canonical Monte Carlo
simulations, we obtained adsorption isotherms for single components and gas mixtures, partial molar heats of adsorption of single components, and the Helmholtz
energy of a hydrate with the gas mixture referred to an empty hydrate.

The results show that the state of the hydrate is determined by the relation between the size of the clathrate cages and the guest molecules. The sI hydrate unit
cell has two types of cages, 6 large and 2 small ones. The small methane molecules can easily fill both the large and the small cages, while the larger and not
spherically symmetric molecules of carbon dioxide prefer the large cages. This determines the two regions of the hydrate state with various loading.

At a small pressure the \co molecules tend to fill the large cages only. Every new adsorbed molecule decreases the Helmholtz energy of the gas mixture in the
hydrate so that the hydrate becomes more stable. This happens until all the large cages are occupied. Then the small cages start to be filled. While the hydrate is
still becoming more stable (the Helmholtz energy is decreasing), the relative increase in stability becomes smaller.

The methane adsorption reveals almost no preference to cage type, so the partial molar heat of adsorption is always decreasing, which means that every new molecule
is more welcome than the previous one. The Helmholtz energy is always decreasing, which means that the fully occupied hydrate is the most stable one comparing to a
partially filled hydrate.

Furthermore, we studied the temperature dependence of the loading, the Helmholtz energy and the partial heat of adsorption. While the adsorption isotherm of
single-component hydrate shifts when the temperature is changed, the dependence of the Helmholtz energy and the partial heat of adsorption on loading is unaltered
by the change of the temperature. This indicates that the adsorption may be considered as ideal, i.e. the interaction between adsorbed molecules in different cages
is low.

The single-component adsorption behavior determines the regime of mixture adsorption. The adsorbed mixture may be regarded as ideal, but only when the large cages
are being filled. At this point there is almost no difference between the behavior of \co and \ch molecules. In contrast, when small cages start to be filled, the
hydrate with gas mixture reveals strong preference to \ch adsorption.

Finally, the diagram obtained for the Helmholtz energy of the gas mixture in hydrate is promising. It may suggest a way to convert \ch hydrates, which are
excessively available at the sea bottoms, to \co hydrates. This could play a role for the question of \co storage, without destroying a hydrate. If one considers
\figr{fig/FPY_CO2+CH4} as a starting point, a possible path corresponds to going from the top left corner (pure \ch hydrate, full loading) to the top right corner
(pure \co hydrate, full loading). In order to keep hydrate stable, i.e. to not use extra energy to destroy it, one has to go along the colored path on
\figr{fig/FPY_CO2+CH4_p}. From the kinetic point of view it may be complicated to perform an exchange of molecules in a hydrate, since, unlike zeolites, a hydrate
does not have empty channels for molecules to go in. In practice, the exchange could be performed in a distorted clathrate structure \cite{Peters2008}. Existence of
such a path must presently be seen as qualitative, as we are lacking information on the thermodynamic properties of the the other phases which can be in equilibrium
with the hydrate phase, cf. \eqr{eq/Intro/01}. Our further efforts are now directed into calculating the phase diagram and the remaining thermodynamic data.

\section{Acknowledgment}
We would like to thank Juan-Manuel Castillo Sanches for help with simulations and Bj{\o}rn Kvamme for fruitful discussions. We are also grateful to the VISTA grant
\#6343.

\clearpage
\clearpage
\clearpage
\clearpage
\clearpage
\clearpage
\clearpage
\clearpage
\clearpage
\clearpage
\clearpage
\clearpage
\clearpage
\begin{figure}[h]
\centering
\subfigure[] %
{\includegraphics[scale=\scaleprofile]{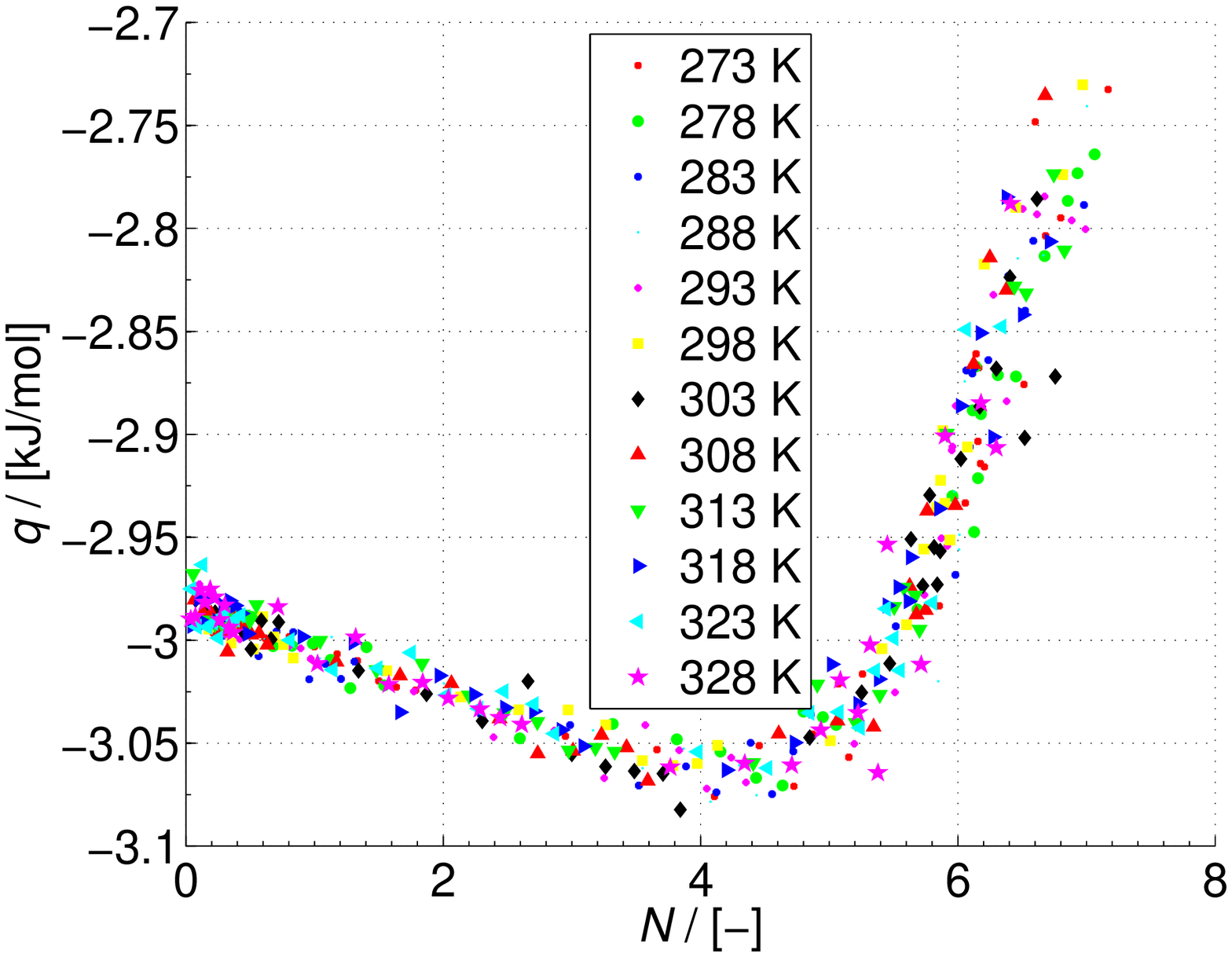}\label{fig/T_HN-CO2} } %
\subfigure[] %
{\includegraphics[scale=\scaleprofile]{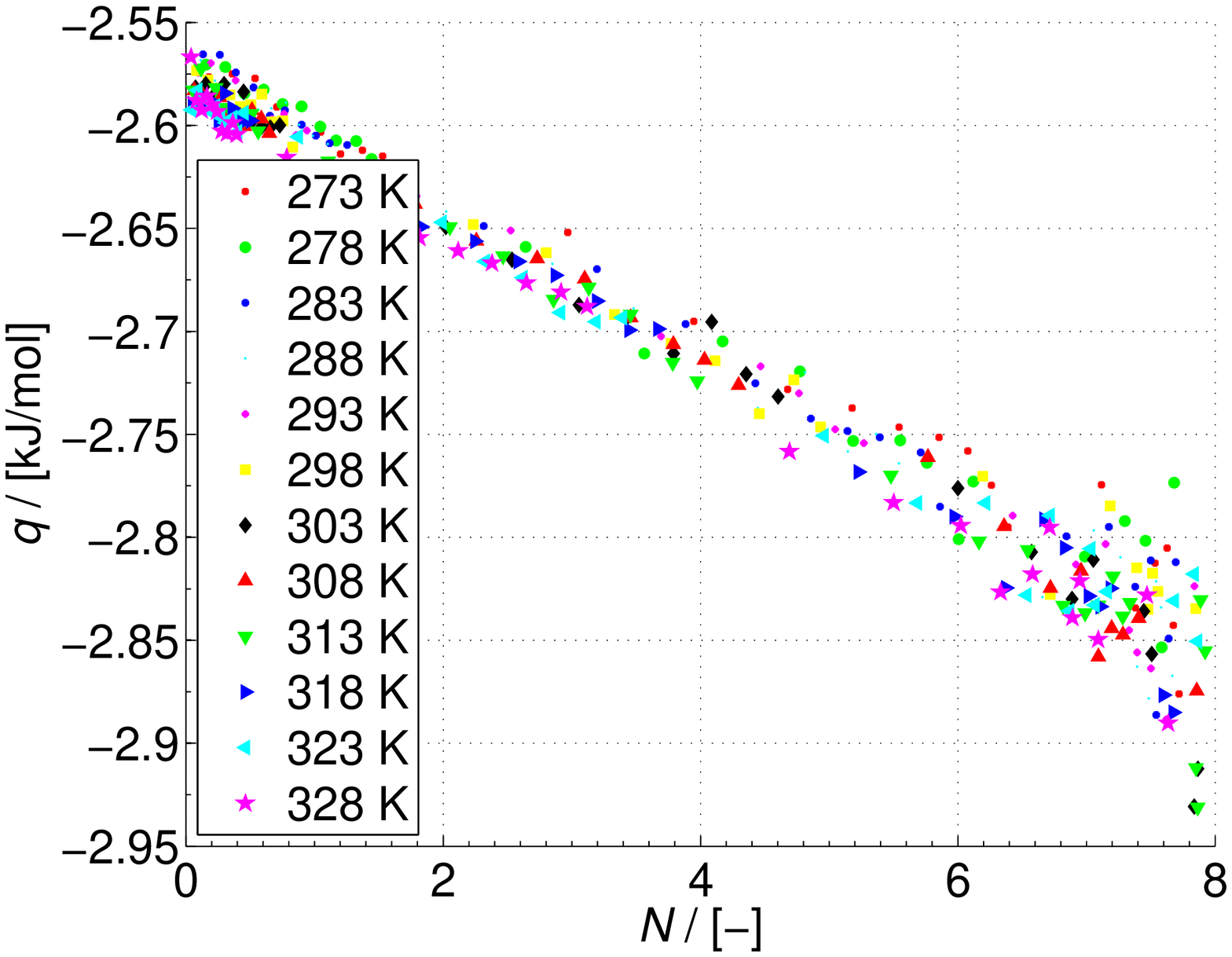} \label{fig/T_HN-CH4} } %
\caption{Partial molar heat of adsorption of a single component hydrate as computed by GCMC simulations at different temperatures as a function of the loading of a
(a) \co (b) \ch hydrate.} \label{fig/T_HN}
\end{figure}
\clearpage

\bibliographystyle{unsrt}

\end{document}